\documentclass[journal]{IEEEtran}
\usepackage{amsmath,amsfonts}
\usepackage{algorithmic}
\usepackage{algorithm}
\usepackage{array}
\usepackage{tabularx}
\usepackage[caption=false,font=normalsize,labelfont=sf,textfont=sf]{}
\usepackage{textcomp}
\usepackage{amsmath}
\usepackage{mathtools}
\usepackage{stfloats}
\usepackage{lipsum}  
\usepackage{url}
\usepackage{verbatim}
\usepackage{caption}
\usepackage{graphicx}
\graphicspath{{photo/}}
\usepackage{cite}
\usepackage{placeins}
\usepackage[font=small,labelfont=bf]{caption}
\usepackage{subcaption}
\begin{document}

\title{Microfluidic Molecular Communication Transmitter Based on Hydrodynamic Gating}
\author{Iman Mokari Bolhassan,~\IEEEmembership{Student Member,~IEEE}, Ali Abdali,~\IEEEmembership{Student Member,~IEEE},
        and \\
        Murat Kuscu,~\IEEEmembership{Member,~IEEE}
       
     \thanks{The authors are with the Nano/Bio/Physical Information and Communications Laboratory (CALICO Lab), Department of Electrical and Electronics Engineering, Koç University, Istanbul, Turkey (e-mail: \{ibolhassan22, aabdali21, mkuscu\}@ku.edu.tr).}
	   \thanks{This work was supported in part by the European
Union’s Horizon 2020 Research and Innovation Programme through the Marie Skłodowska-Curie Individual Fellowship under Grant Agreement 101028935, and by The Scientific and Technological Research Council of Turkey (TUBITAK) under Grant \#120E301.}}



\maketitle

\begin{abstract}

Molecular Communications (MC) is a bio-inspired paradigm for transmitting information using chemical signals, which can enable novel applications at the junction of biotechnology, nanotechnology, and information and communication technologies. However, designing efficient and reliable MC systems poses significant challenges due to the complex nature of the physical channel and the limitations of the micro/nanoscale transmitter and receiver devices. In this paper, we propose a practical microfluidic transmitter architecture for MC based on \emph{hydrodynamic gating}, a widely utilized technique for generating chemical waveforms in microfluidic channels with high spatiotemporal resolution. We develop an approximate analytical model that can capture the fundamental characteristics of the generated molecular pulses, such as pulse width, pulse amplitude, and pulse delay, as functions of main system parameters, such as flow velocity and gating duration. We validate the accuracy of our model by comparing it with finite element simulations using COMSOL Multiphysics under various system settings. Our analytical model can enable the optimization of microfluidic transmitters for MC applications in terms of minimizing intersymbol interference and maximizing data transmission rate.

\end{abstract}


\section{Introduction}
\IEEEPARstart{T}{he} Internet of Bio-Nano Things (IoBNT) is an emerging networking paradigm that involves the communication and cooperation of synthetic and natural biological entities with diverse micro/nanoscale devices to enable novel healthcare and environmental applications \cite{akyildiz2015internet, kuscu2021internet, akan2023internet}. Molecular Communications (MC), using molecules to transfer information, is regarded as the most viable communication technology to enable IoBNT, offering significant advantages in complex and challenging environments, such as inside human body, which are not conducive to efficient transmission of electromagnetic signals \cite{kuscu2022detection, akan2016fundamentals, akyildiz2011nanonetworks}. Despite significant theoretical research on MC, many practical aspects have not yet been thoroughly addressed, leading to a gap between theory and practice. However, experimental studies have recently gained momentum with through the utilization of microfluidic technologies that replicate the geometry and flow conditions of typical biological environments in MC applications \cite{kuscu2021fabrication, 10102773}. The use of microfluidic testbeds provides the advantage of precise control over fluidic MC channels regarding geometry and flow profile, straightforward fabrication with soft lithography, and easy integration of micro/nanoscale MC transceiver components, such as two-dimensional bioFET-based MC receivers \cite{kuscu2021fabrication}. However, achieving control over the spatiotemporal distribution of information molecules within the microfluidic channels with high precision is a standing issue that leads to complexities in experimental testing and refining the developed MC modulation and detection methods \cite{zadeh2023microfluidic}.

Transmitter is a fundamental component in any communication system, including microfluidic MC systems. Its role is to encode information by manipulating the physical properties of molecules, such as concentration, type, ratio, order, or timing of release, and transmit the modulated molecular signals into the medium \cite{kuscu2019transmitter}. Various MC transmitter designs have been proposed, many of which were recently reviewed in \cite{civas2022molecular}. Particular examples compatible with microfluidic testbeds were presented by Bi et al. in \cite{bi2020chemical, bi2022microfluidic}, that leverage flow- and geometry-controlled chemical reactions to generate predefined molecular concentration pulses in response to rectangular triggering signals. However, the method involves chemical reactions, increasing the number of control parameters and thus amplifying the complexity of design and the input-output relationship. Moreover, the slow kinetics of reactions restrict the bandwidth of molecular signals that can be generated. In other microfluidic testbeds, information molecule concentration modulation within a microfluidic channel was achieved via flow velocity modulation, typically in one of the branches of Y-junction inlets \cite{angerbauer2023salinity, kuscu2021fabrication}. This subsequently results in mixing the solutions carrying information molecules with buffer solutions, yielding a time-varying concentration profile in the channel. However, this approach results also in significant modulation of the flow profile (e.g., flow velocity) in the channel and prevents precise control over the temporal profile of concentration signals due to the turbulent nature of the mixing. To overcome these limitations, in this paper, we propose a novel microfluidic MC transmitter design based on hydrodynamic gating, which offers high spatiotemporal control over the generated molecular concentration pulses that encode information.

Hydrodynamic gating involves modulating the flow velocity and direction to control the injection rate of molecules and shape their concentration waveforms as they propagate through the channel \cite{chen2013analysis, guo2020multichannel}. Known for its high precision and biocompatibility, hydrodynamic gating is well-suited for use in biological research, particularly for modulating the molecular composition of extracellular matrices in cultured cells \cite{chen2020review}. Moreover, its implementation and operation are straightforward in conventional polydimethylsiloxane (PDMS)-based microfluidic systems  \cite{guo2021time}. With a relatively small number of design parameters that can be fine-tuned using standard soft-lithography techniques and pressure control devices, hydrodynamic gating has proven a reproducible method, which is essential for reliable and consistent experiments \cite{zadeh2023microfluidic}. 

The application of hydrodynamic gating for microfluidic MC transmitter action involves the release of information molecules, i.e., ligands, with particular temporal concentration waveforms, and subsequent transportation of the generated concentration signals through convection and diffusion in a microfluidic channel. However, this process is characterized by significant nonlinearity and time-variance, making it impossible to obtain an exact analytical solution for the spatiotemporal evolution of the concentration profile of transmitted ligands in the microfluidic channel. Consequently, the use of computationally-intensive numerical techniques, such as finite element analysis (FEA), becomes imperative. In order to characterize and optimize the system from a communication theoretical perspective, in this paper, we develop a parametric analytical model which approximates the input-output relationship between control parameters, such as flow velocity at the inlets, gating duration, and the fundamental characteristics of the pulses generated by the microfluidic channel, such as pulse amplitude, width, and delay. The accuracy of the analytical model is evaluated by comparing the analytical results and numerical solutions obtained from finite element simulations in COMSOL Multiphysics.

The structure of this paper is organized as follows: Section \ref{sec:hydrodynamic gating model} details the operation of the proposed hydrodynamic gating-based MC transmitter. In Section \ref{sec:exact model}, We provide the exact end-to-end mathematical model of the MC transmitter, incorporating nonlinear partial differential equations. Section \ref{sec:analytical model} presents our approximate analytical model composed of two main components: i) pulse generation compartment, and ii) pulse propagation compartment. In this section, we provide the derived analytical expressions for the three main characteristics of the propagating concentration pulses, e.g., pulse amplitude, pulse width, and pulse delay. Furthermore, we optimize our model in terms of accuracy using a genetic algorithm implemented in MATLAB. Section \ref{sec:successive pulse Transmitter}, we extend the presented analytical model to the successive-pulse transmission mode. In section \ref{section_results}, we evaluate the accuracy of our model by comparing its results with simulations and investigate the impact of input parameters on the characteristics of the output signal. Finally, the paper concludes in the last section.

\section{Hydrodynamic Gating For Molecular Communication Transmitter}
\label{sec:hydrodynamic gating model}

\begin{figure}[t!]
    \centering
    \includegraphics[width=0.48\textwidth]{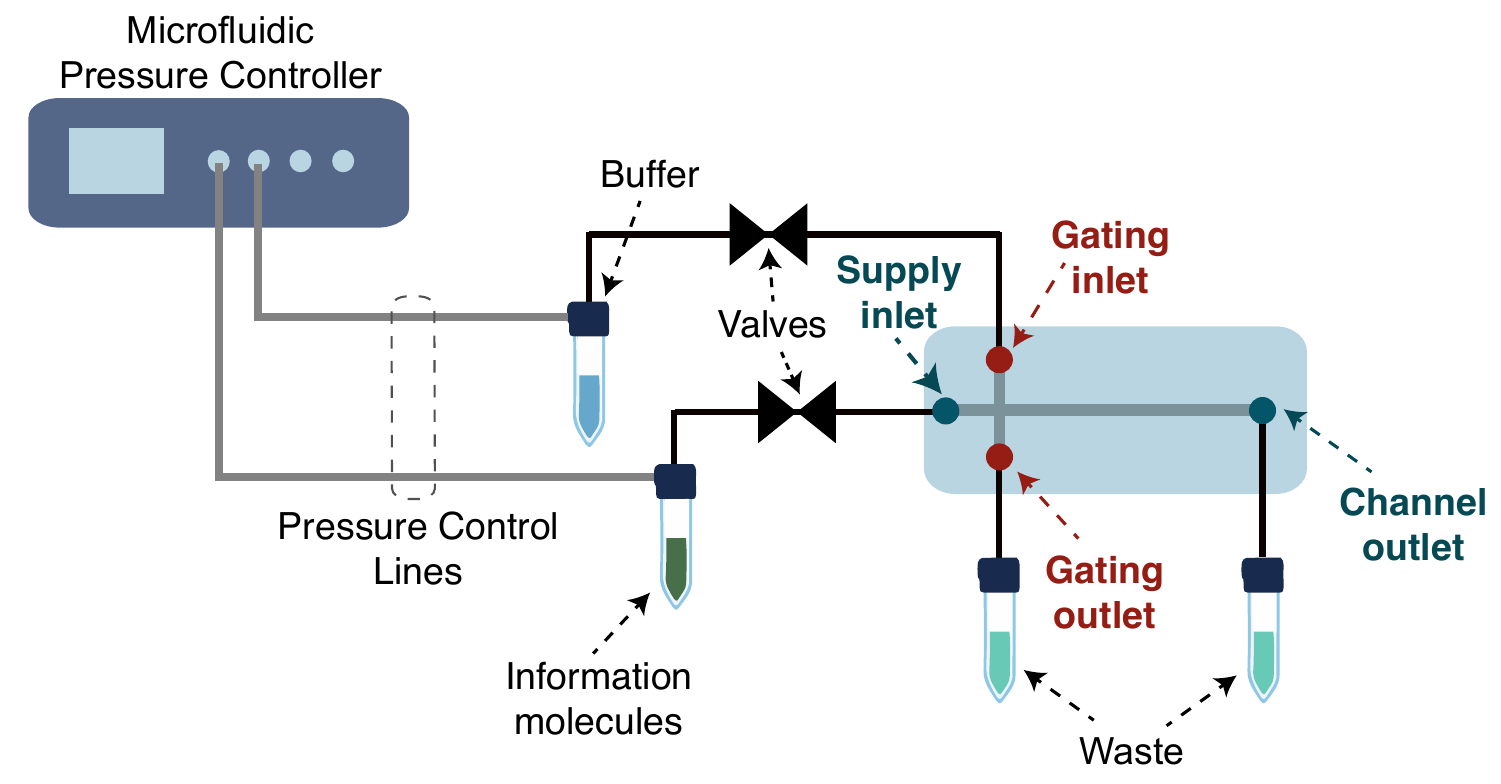}
        \caption{Schematic representation of microfluidic MC transmitter system setup based on hydrodynamic gating technique.}
    \label{fig:hydrodynamic}
\end{figure}

An example microfluidic setup for implementing a hydrodynamic gating technique in MC transmitter is depicted in Fig. \ref{fig:hydrodynamic}. It consists of a cross-shaped microfluidic channel, which can be fabricated using soft lithography techniques and polydimethylsiloxane (PDMS) polymer. Pressure control at the microchannel inlets can be achieved using custom-built or commercial pressure control systems such as programmable syringes. In this setup, hydrodynamic gating is achieved by switching on and off the laminar flow, which comes from the supply inlet and transports the molecules of interest, through the on-off modulation of the laminar flow from the gating inlet. When the gating is turned on, the flow from the gating inlet generates a hydrodynamic force across the cross-section that is sufficient to cut the flow from the supply inlet. As a result, the molecules transported through the continuous flow from the supply inlet are unable to enter the propagation channel and are directed toward the gating outlet. On the other hand, when gating is turned off for a short duration, the vertical hydrodynamic force is removed, allowing the molecules coming from the supply inlet to be injected into the propagation channel. The gating duration is critical in determining the width of the pulses propagating inside the channel. Other design parameters that play essential roles in shaping the generated molecular waveforms include the absolute and relative flow velocities or pressures applied at the supply and gating inlets, as well as the concentration of molecules transported from the supply inlet.

\begin{figure}
    \centering
    \includegraphics[width=0.46\textwidth]{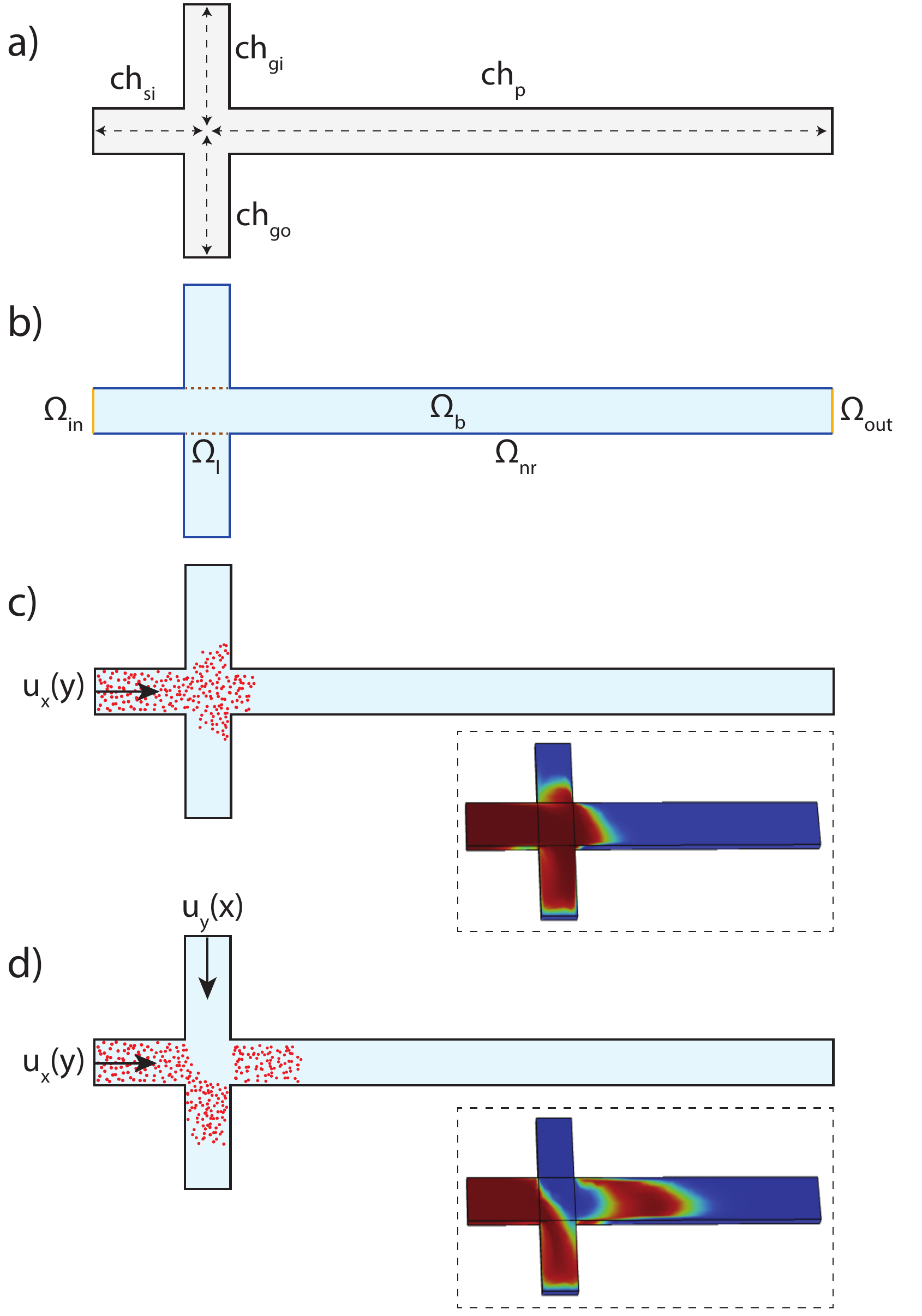}
    \caption{Microfluidic MC channel with hydrodynamic gating technique.}
    \label{fig:exact}
\end{figure}

\begin{figure*}[t!]
     \centering
     \begin{subfigure}[b]{0.32\textwidth}
         \centering
          \includegraphics[width=\textwidth]{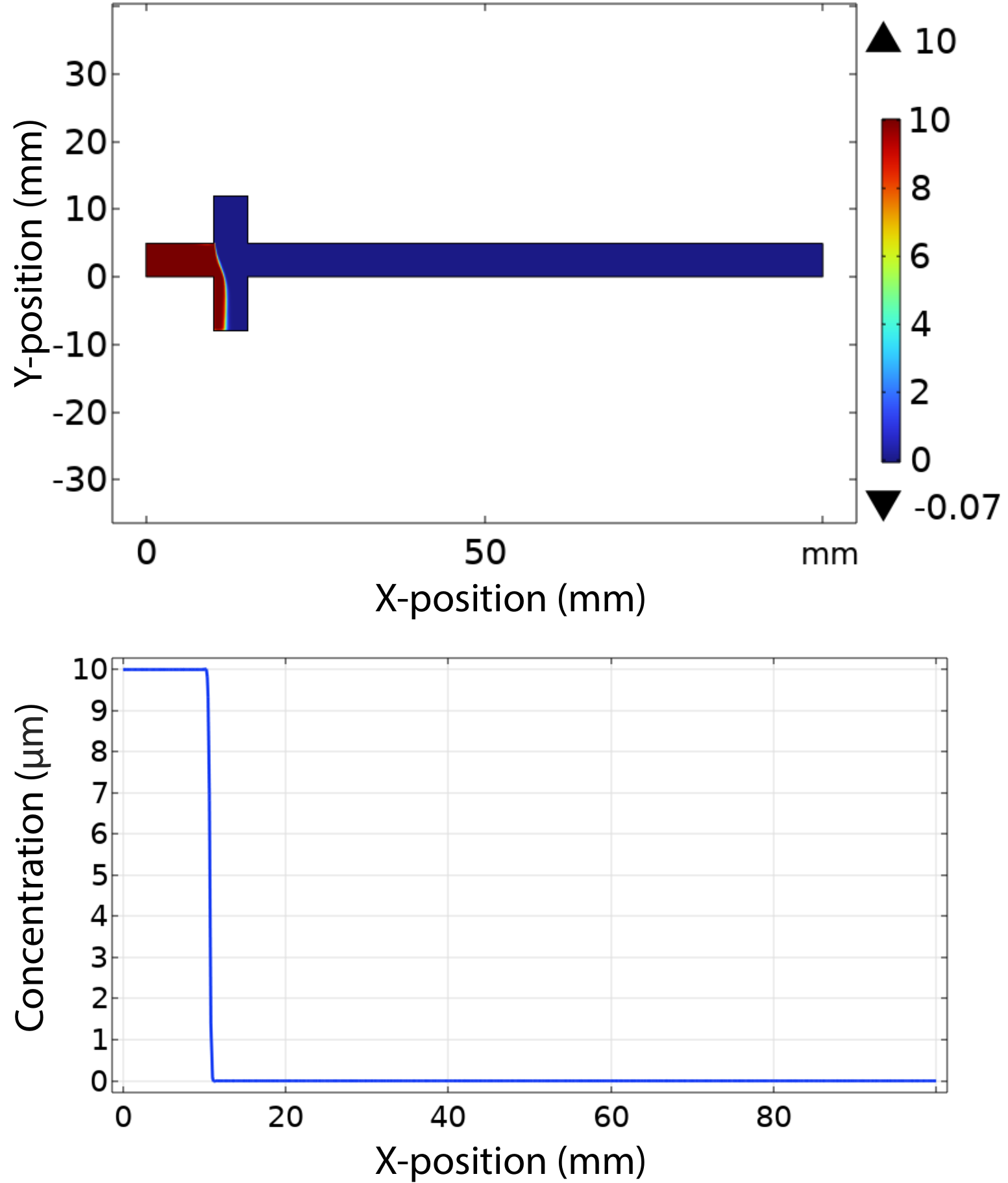}
         \caption{}
         \label{}
    \end{subfigure}
    \hfill
         \begin{subfigure}[b]{0.32\textwidth}
         \centering
          \includegraphics[width=\textwidth]{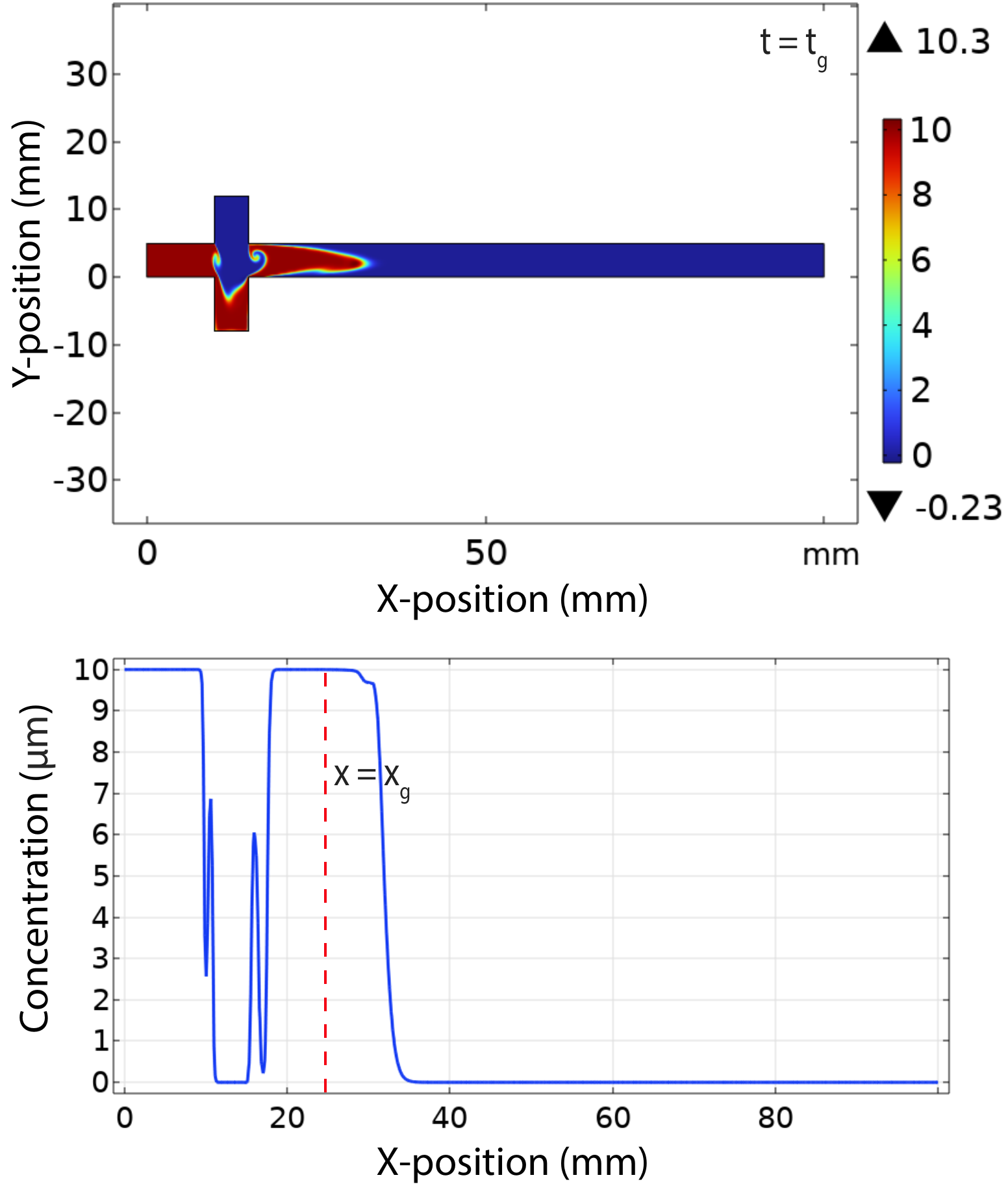}
         \caption{}
         \label{}
    \end{subfigure}
    \hfill
         \begin{subfigure}[b]{0.32\textwidth}
         \centering
          \includegraphics[width=\textwidth]{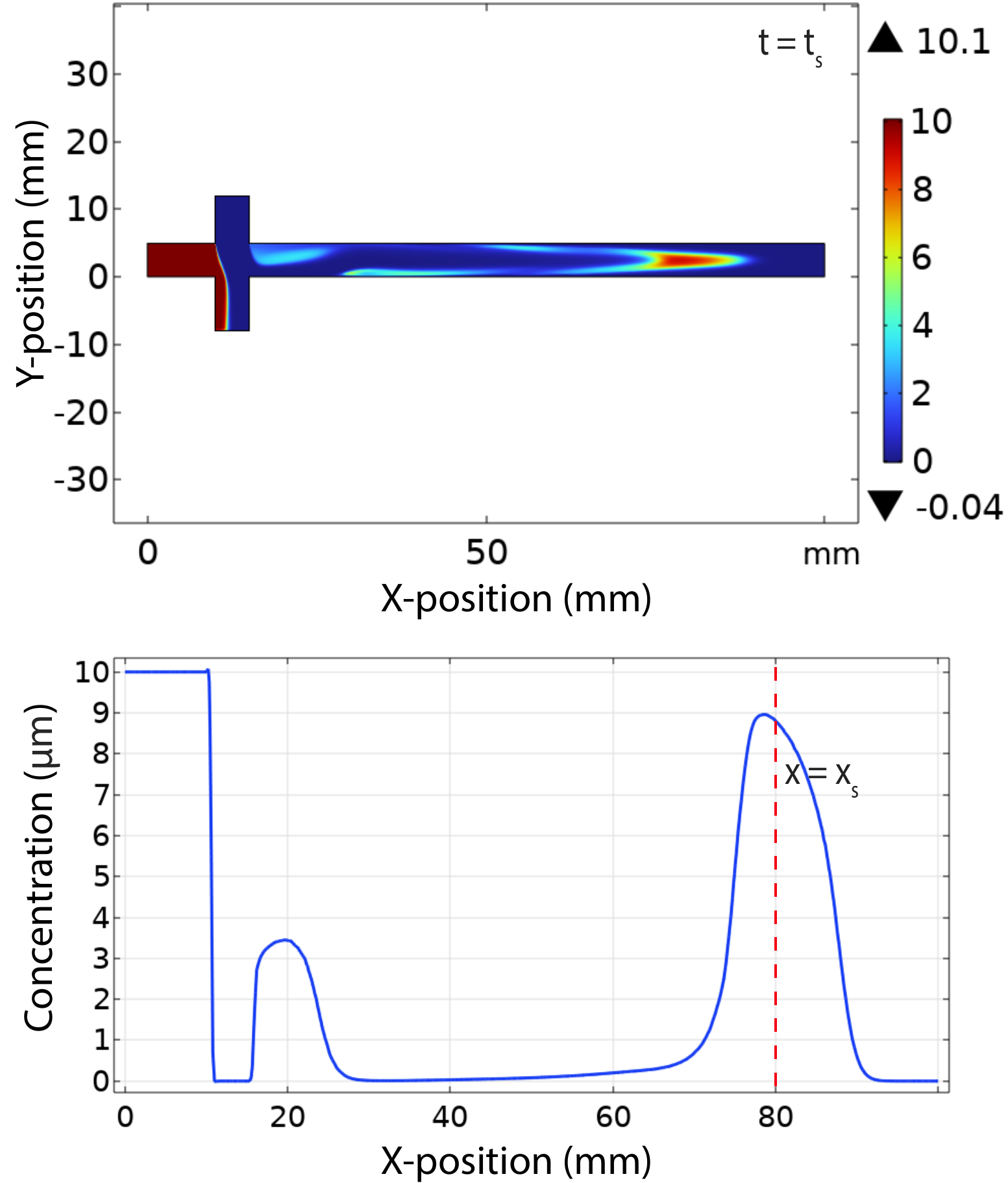}
         \caption{}
         \label{}
    \end{subfigure}
     \caption{An illustration of a concentration pulse sampled at three different time instances: (a) t = 10.00s, (b) t$_\mathrm{g}$ = 12.18s, and (c) t$_\mathrm{s}$ = 15.19s, corresponding to different states of the hydrodynamic gating process. Part a and c depict the gating state, and Part b depicts the injection state, simulated with COMSOL Multiphysics.}
         \label{fig:hydrodynamic-single pulse}
    \end{figure*}

To illustrate the microfluidic transmission of molecular concentration pulses via hydrodynamic gating, we perform a set of finite-element simulations using COMSOL Multiphysics (The default values for the system parameters used in these simulations are listed in Table \ref{table:inputs}). In these simulations, we use a two-dimensional (2D) model of microfluidic hydrodynamic gating system, as illustrated in Fig. \ref{fig:exact}(a). This model is able to capture hydrodynamic gating effect and the subsequent propagation of molecules inside the microfluidic channel through convection and diffusion. 
The finite-element simulation results for the generation of a molecular concentration pulse are provided in Fig. \ref{fig:hydrodynamic-single pulse}, which demonstrates the molecular concentration profile in the microfluidic channel during three fundamental steps of the hydrodynamic gating process: (a) the first gating state, (b) The injection state, and (c) the second gating state. the first raw of the figure shows the 2D concentration profile of molecules sampled at specific times, while the second raw plots the concentration of molecules sampled across the entire microfluidic channel from the supply inlet to the channel outlet using a concentration probe at the middle of the channel.

\begin{table}[t!]
\centering
\caption{Default values of simulation parameters}
\label{table:inputs}
\scalebox{0.8}{
\begin{tabular}{m{8cm}|m{2 cm}}
\hline
\hline
Applied initial concentration of ligands  from the supply inlet ($c_\mathrm{s}$) &  $10^{-2}$ mol/m$^3$\\ 
\hline
Applied average velocity from the supply inlet ($u_\mathrm{s}$) & $10^{-2}$ m/s    \\
\hline
Applied average velocity from the gating inlet ($u_\mathrm{g}$) & $5\times10^{-2}$ m/s \\
\hline
 Diffusion coefficient of ligands ($D$) & $10^{-10}$ m$^2$/s \\
 \hline
The relative ratio between applied flow from inlets  ($r_\mathrm{u}$ = $u_\mathrm{g}$ / $u_\mathrm{s}$) & $5$  \\
\hline
The duration of the removal of the gating inlet ($T_\mathrm{g}$) & $2$ s\\
\hline
Supply inlet channel length  ($l_\mathrm{s}$) & $12.5$ mm  \\
\hline
Gating inlet channel length ($l_\mathrm{gi}$) & $12.5$ mm  \\
\hline
Gating outlet channel length ($l_{go}$) & $12.5$ mm  \\
\hline
Propagation channel length ($l_{p}$) & $87.5$ mm  \\
\hline
Microfluidic channel length along the x-axis ($l_\mathrm{ch}=l_\mathrm{s}+l_{p}$) & $100$ mm  \\
\hline
Microfluidic channel width ($w_{ch}$) & $5$ mm  \\
\hline
Time interval between two consecutive gatings ($T_p$) & $2$ s \\
 \hline
Pulse sampling point along the x-axis ($x_\mathrm{s}$) & $80$ mm \\
\hline
 Pulse generation point along the x-axis ($x_\mathrm{g}$) & $25$ mm \\
\hline
Pulse generation time  ($t_\mathrm{g}$)  & $12.18$ s \\
\hline
Pulse sampling time ($t_\mathrm{s}$) &  $15.19$ s \\
\hline

\hline

\end{tabular}
}
\end{table}

During the first gating state (a), a vertical flow from the gating inlet diverts all molecules from the supply inlet to the gating outlet, preventing them from entering the microfluidic channel through the cross-section. During the injection state, the gating flow is turned off for a brief period, allowing a small number of molecules to enter the microfluidic channel without being diverted. When the gating flow is turned back on, the system returns to its initial state, halting the transport of molecules into the microfluidic channel. As a result, a short concentration pulse of molecules is generated and travels towards the channel outlet. As the concentration pulse moves along the microfluidic channel, it experiences dispersion due to both diffusional and convective transport. The interplay between diffusion and convection determines pulse dispersion, which is a critical metric related to inter-symbol interference (ISI) in MC applications.

In this study, our goal is to develop an analytical model that approximates the waveform of the generated and propagated concentration pulses inside the microfluidic channel. This model establishes a mathematical relationship between the system inputs and the three main parameters of the transmitted pulse: pulse amplitude, pulse width, and pulse delay. The approximate model can facilitate the programmability and optimal use of pulse characteristics for encoding information in microfluidic MC systems.

\section{Exact End-to-End Model}
\label{sec:exact model}
In this section, we present the end-to-end model of the microfluidic communication channel, representing it as a set of partial differential equations.  We utilize a 2D model, as molecular transport can be effectively modeled in 2D, particularly when there is an interplay between convection and diffusion in the system. In this model, both diffusion and convection contribute to the transportation of ligands from the supply inlet, denoted as $ch_{\mathrm{si}}$, to the propagation channel, represented by $ch_{\mathrm{p}}$, which together form the main channel.

As illustrated in Fig. \ref{fig:exact}(b), we have defined three orthogonal domains: (I) bulk domains $\Omega_{\mathrm{b}}$, where the convection and diffusion of ligands take place, (II) leakage domains $\Omega_{\mathrm{l}}$, where a portion of ligands concentration supplied from inlet leaks towards the gating inlet $ch_{\mathrm{gi}}$ and gating outlet $ch_{\mathrm{go}}$ as shown in Fig. \ref{fig:exact}(c), and (III) non-reacting surface domains $\Omega_{\mathrm{nr}}$, which define the non-reactive walls of the microfluidic channel. When the gating is activated, the flow from the gating inlet interrupts the transport of ligands from the supply inlet to the propagation channel, as depicted in Fig. \ref{fig:exact}(d). We also define a variable $c=c(x,y,t)$ representing ligand concentration in space and time.

In the absence of gating, convection-diffusion equation governs ligand propagation in a 2D microfluidic channel:
\begin{equation}
     \frac{\partial c}{\partial t} = D\nabla^{2} c- u_{\mathrm{x}}(y) \nabla c + R,
\end{equation}
where $c$ is the concentration of ligands, $D$ is the diffusion coefficient, $\nabla^2=\frac{\partial^2}{\partial x^2}+\frac{\partial^2}{\partial y^2}$ represents the 2D Laplace operator, $ u_{\mathrm{x}}(y)$ represents the velocity field as a function of distance to the surrounding walls, and $R$ describes chemical reactions between ligands. However, in this model, we assume $R=0$ because of the negligible chemical reactions between ligands and having non-reacting walls of the channels. In the absence of gating, we assume a completely developed laminar flow in the main channel ($ch_{\mathrm{si}}+ch_{\mathrm{p}}$), resulting in a parabolic flow velocity profile as
\begin{align}
    u_{\mathrm{x}}(y)=4u(y/h)(1-y/h), \hspace{5 mm} (x,y)\in \Omega_{\mathrm{b}}.
\end{align}
When the gating is turned on, an additional parabolic flow $u_{\mathrm{y}}(x)$ emerges from the gating inlet to the gating outlet, intersecting perpendicularly with the existing flow $u_{\mathrm{x}}(y)$. The combined flow at the intersection point can be expressed as $\lvert u_{\mathrm{xy}}(x,y)\rvert = \sqrt{u_{\mathrm{x}}(y)^2+u_{\mathrm{y}}(x)^2}$, and its direction is given by $\theta=-\tan^{-1}(\lvert\frac{u_{\mathrm{y}}(x)}{u_{\mathrm{x}}(y)}\rvert)$. 

No-flux boundary condition is assigned to the non-reacting walls of the channel, i.e.,
\begin{align}
    \frac{\partial c}{\partial y}=0 , \hspace{5mm} y  \in \Omega_{\mathrm{nr}}.
\end{align}
Flux condition at the intersection of the main channel and the gating channel, corresponding to the leakage boundary, is expressed as 
\begin{align}
    -D\frac{\partial c}{\partial y}=\mathcal{L}(A_\mathrm{g},c), \hspace{5mm} y  \in \Omega_{\mathrm{l}},
\end{align}
where $A_\mathrm{g}=w_{\mathrm{g}}\times h_{\mathrm{g}}$ represent the cross-sectional area of the gating channel, with $w_{\mathrm{g}}$ and $h_{\mathrm{g}}$ denoting the width and height of the gating channel, respectively, and $\mathcal{L}$ represents the leakage flux. We define the inlet and outlet boundary conditions in terms of ligand concentration as 
\begin{align}
    c(x=0,y,t)= c_{\mathrm{in}}(t),\\
    \frac{\partial c(x=L,y,t)}{\partial x}=0,
\end{align}
where $c_{\mathrm{in}}(t)$ is the transmitted signal. Finally, the initial conditions for the system are defined as 
\begin{align}
    \mathcal{L}(x,0)&=0, \hspace{5mm} (x,y=0,w_\mathrm{g})\in\Omega_{\mathrm{l}}.\\
    c(x,y,0)&= 0,\hspace{5mm} (x,y)\in\Omega_{\mathrm{b}}.
\end{align}

Due to the complexity of the exact system model presented above, it is not analytically tractable. As a result, numerical methods, such as finite element simulations, are required to determine the concentration of ligands in time and space along the propagation channel. Therefore, in the following section, we present an approximate analytical method to obtain ligand concentration profile in the propagation channel.

\section{Approximate Analytical Model}
\label{sec:analytical model}

\begin{figure}[t]
      \centering
       \includegraphics[width=0.5\textwidth]{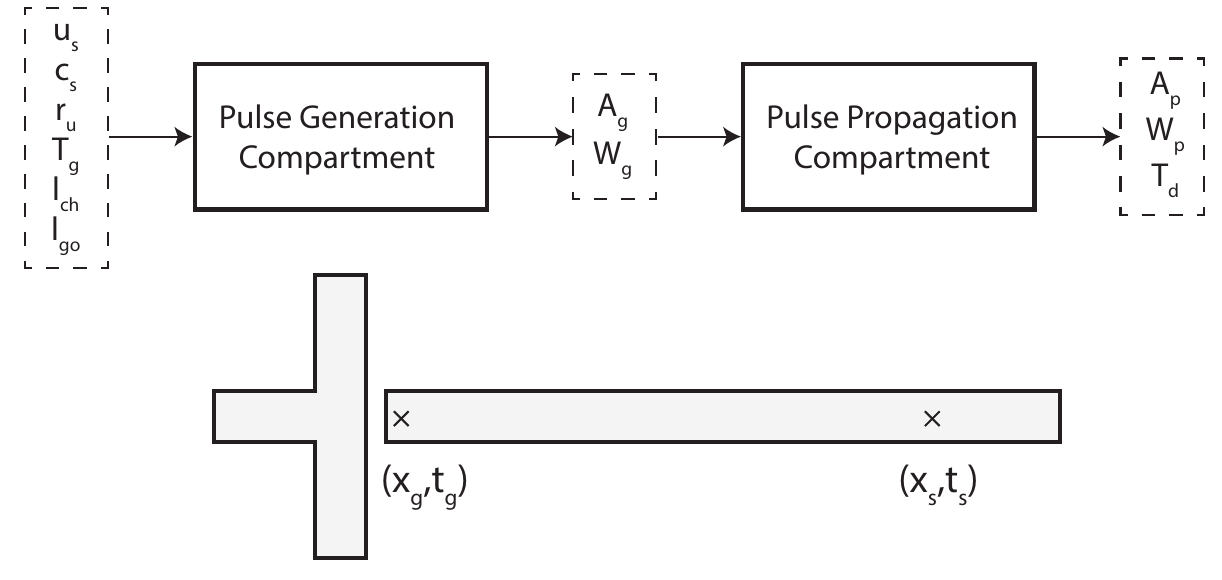}
       \caption {Block diagram and schematic of the hydrodynamic gating system divided into two compartments.}
          \label{fig:strategy}
\end{figure}

Characteristics of the transmitted concentration pulse are revealed by finite element simulations as depicted in Fig. \ref{fig:hydrodynamic-single pulse}. The simulations demonstrate the pulse at the pulse generation point ($x_\mathrm{g}$) during the pulse generation time ($t_\mathrm{s}$), as seen in Fig. \ref{fig:hydrodynamic-single pulse}(b). Additionally, Fig. \ref{fig:hydrodynamic-single pulse}(c) displays the concentration pulse at an arbitrary sampling point ($x_\mathrm{s}$) along the propagation channel ($ch_\mathrm{p}$) during the sampling time ($t_\mathrm{s}$). The one-dimensional propagation profiles of this pulse indicate that the Gaussian function well approximates the waveform of the concentration pulse propagating in the microfluidic channel, and it can be mathematically represented as follows:
\begin{align}
\label{Gaussian1}
C(x)|_{x=x_\mathrm{s}} = A_\mathrm{p} \exp \left(-\frac{(x - x_\mathrm{s})^2}{2 \sigma^2} \right), 
\end{align}
where $A_\mathrm{p}$ corresponds to the pulse amplitude, representing the peak value of the concentration pulse at $x_\mathrm{s}$, and $\sigma^2$ denotes the variance of the pulse. By utilizing the full width at half maximum (FWHM) method, we can compute the pulse width of the concentration pulse at $x_\mathrm{s}$ as $W_\mathrm{p} = 2.3548\sigma$. Consequently, \eqref{Gaussian1} can be reformulated as follows:
\begin{equation}
\label{Gaussian2}
\left.C(x)\right|_{x=x_\mathrm{s}} = A_\mathrm{p} \exp \left(-\frac{(x - x_\mathrm{s})^2}{0.36 (W_\mathrm{p})^2} \right). 
\end{equation}
As a result, the spatiotemporal distribution of the molecules within the microfluidic channel can be concisely represented as follows:
\begin{equation}
C(x_\mathrm{s},t_\mathrm{s}) = f(A_\mathrm{p},W_\mathrm{p}, T_\mathrm{d}),
\end{equation}
where $T_\mathrm{d}$ is pulse delay representing the time duration for the concentration pulse to travel from the pulse generation point to the pulse sampling point, is a crucial parameter that will allow us to characterize and optimize this system from the communication theoretical perspective. To mathematically model this intricate and non-linear system, our methodology is to conceptualize it as consisting of two interconnected compartments: \emph{pulse generation compartment} and \emph{pulse propagation compartment}. This envisioning, as depicted in Fig. \ref{fig:strategy}, enables us to analyze the system in two distinct parts, thereby significantly reducing its complexity. The pulse generation compartment is responsible for generating the pulse, while the propagation compartment represents the input-output relationship for the propagation phase of the pulse within the propagation channel ($ch_\mathrm{p}$). The output of the pulse generation compartment is input to the pulse propagation compartment.

\begin{figure*}
     \centering
     \begin{subfigure}[b]{0.40\textwidth}
         \centering
          \includegraphics[width=\textwidth]{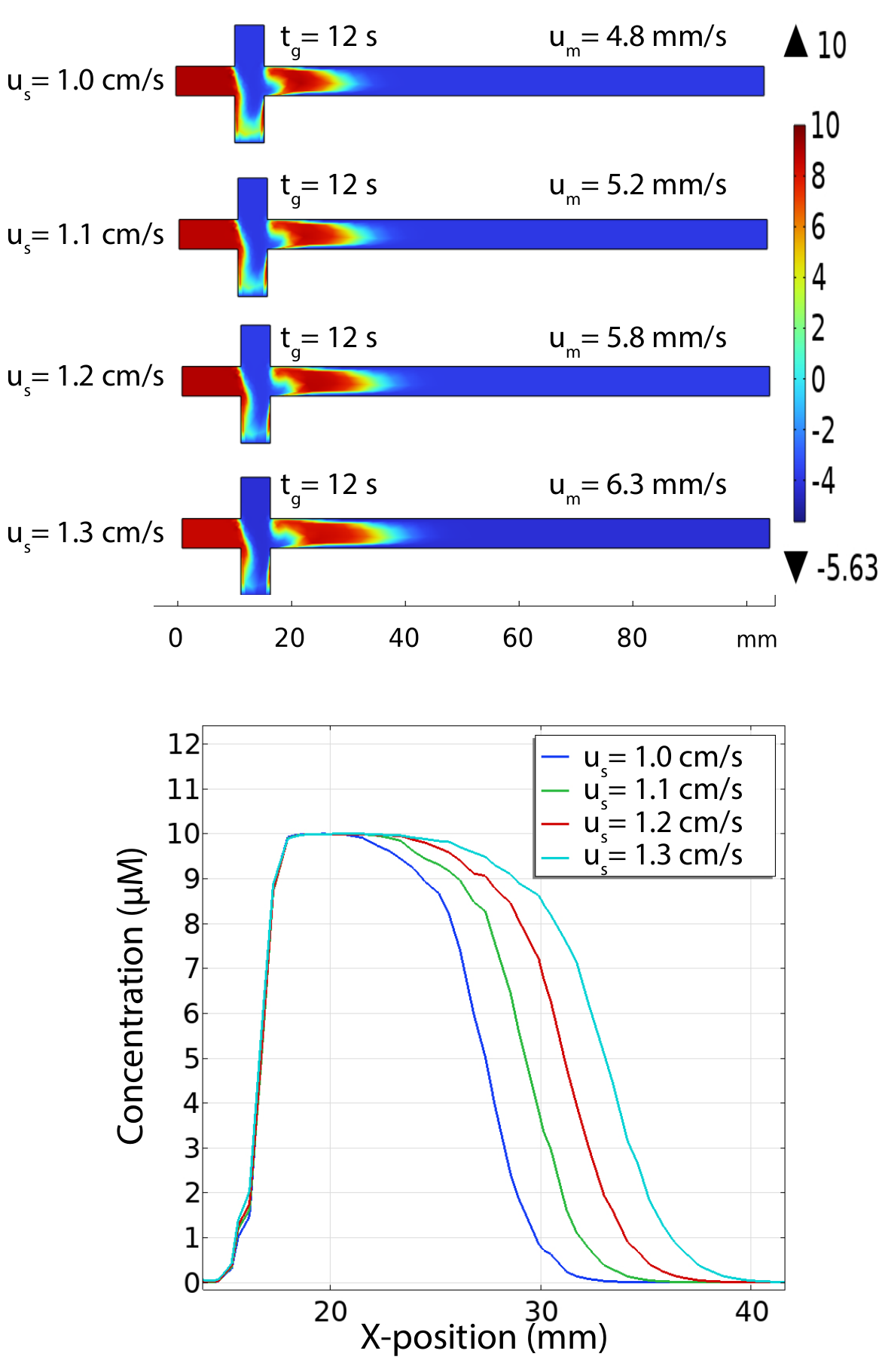}
         \caption{}
         \label{}
    \end{subfigure}
    \hspace{3mm}
         \begin{subfigure}[b]{0.40\textwidth}
         \centering
          \includegraphics[width=\textwidth]{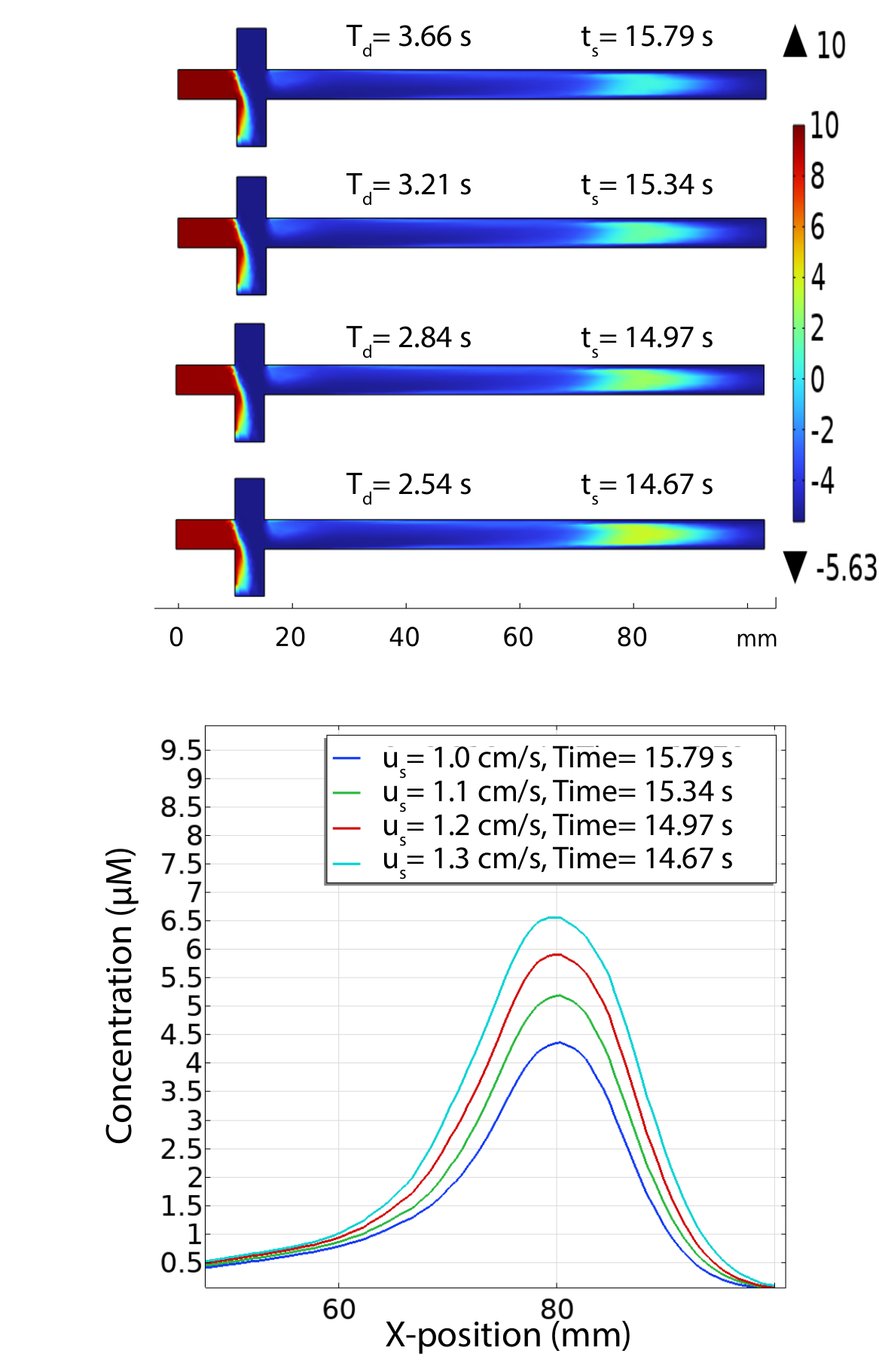}
         \caption{}
         \label{}
    \end{subfigure}
         \caption{The effect of varying supply inlet velocity $u_\mathrm{s}$ on the characteristics of the a) generated pulse, and b) propagated pulse.}
         \label{fig:effect of wg}
\end{figure*}

To validate the efficacy of our selected methodology, finite element simulations were conducted using COMSOL Multiphysics. The outcomes of these simulations are demonstrated in Fig. \ref{fig:effect of wg}, revealing the effect of the various input parameter on both the generated pulse width and the main characteristics of the propagating pulse at the sampling point.
 
\subsection{Pulse Generation Compartment}
To establish the mathematical model governing the pulse amplitude and pulse width of the generated pulse in the output of the pulse generation compartment ($A_\mathrm{g}$ and $W_\mathrm{g}$, respectively), we draw inspiration from a physical phenomenon observed in the model system at the pulse generation time (which was discussed in Section \ref{sec:exact model}). The concentration of information molecules, matching the initially applied concentration from the supply inlet ($c_\mathrm{s}$), enters the propagation channel through the removal of the applied laminar flow from the gating inlet ($u_\mathrm{g}$) for a gating time duration ($T_\mathrm{g}$). The pulse width of the generated pulse, correlating with the number of entered molecules, depends on the gating time duration and the flow velocity in the propagation channel ($u_\mathrm{m}$). An increase in $T_\mathrm{g}$ leads to a higher number of molecules with the same concentration entering the propagation channel, resulting in an increase in $W_\mathrm{g}$. Additionally, a higher flow velocity along with channel $ch_{\mathrm{p}}$ facilitates the entrance of more information molecules into the propagation channel through the convection phenomenon. With inspiration from these two phenomena observed in the model system, we can express the following equations for the output of the pulse generation compartment of this system: 
\begin{equation}
\label{AP1}
A_\mathrm{g} = c_\mathrm{s}\\,
\end{equation}
\begin{equation}\label{wp1_tg__vm}
W_\mathrm{g}= k_\mathrm{g}  T_\mathrm{g}  u_\mathrm{m},
\end{equation}
where $k_\mathrm{g}$ is a fitting parameter that will be determined at the end of this section (the optimization part). This will be accomplished using the genetic algorithm optimization technique in MATLAB. The objective of this optimization is to minimize the error between our analytical model and the corresponding simulation model. The value of $u_\mathrm{m}$ is influenced by the geometry of the microfluidic system and the amplitude of flow velocities applied to the inlets. In order to accurately approximate $u_\mathrm{m}$ in this complex model, it is crucial to conduct a thorough analysis of the model from the perspective of microfluidics. The utilization of the electric circuit analogy of the model is the most viable and effective approach for this purpose\cite{oh2012design}.

\subsubsection{\textbf{Electric Circuit Analogy of the Model}}

\begin{figure}[t!]
\centering
\includegraphics[width=0.45\textwidth]{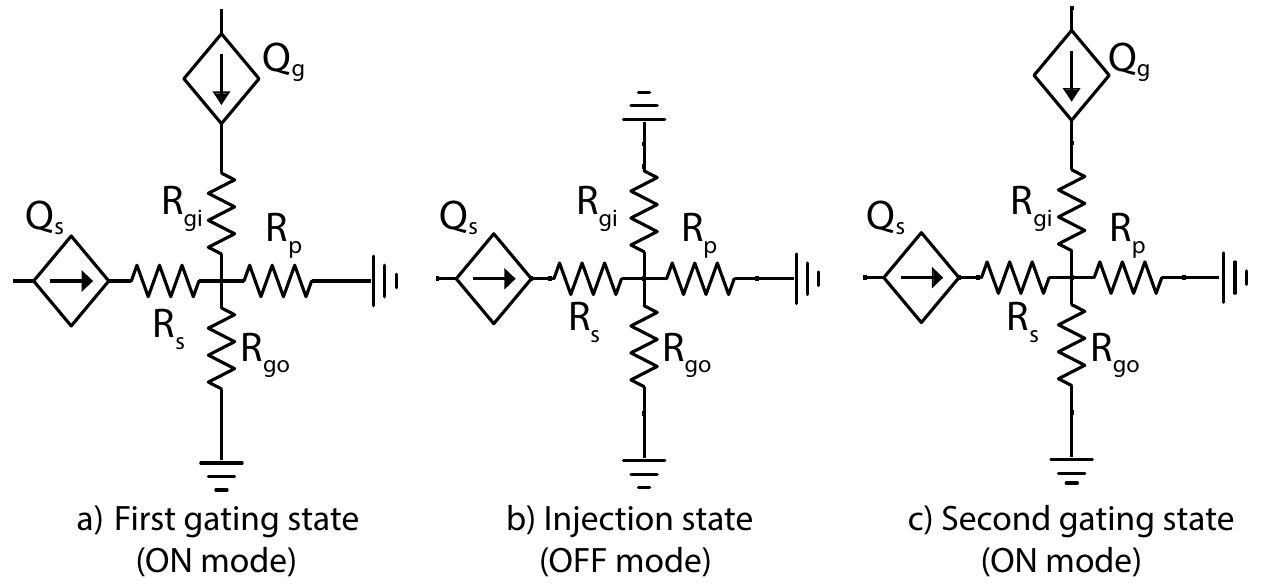}
\caption{Schematic of the circuit-equivalent model of the system at three different states: a) ON mode - First Gating State, b) OFF mode - Injection State, c) ON mode - Second Gating State.}
    \label{fig: Hydrodynamic electrical equivalence}
\end{figure}

The proposed microfluidic model can be equivalently represented as an electrical circuit in different states of this system, as illustrated in Fig. \ref{fig: Hydrodynamic electrical equivalence}. In this analogy, $R_\mathrm{s}$, $R_\mathrm{gi}$, $R_\mathrm{go}$, and $R_\mathrm{p}$ represent the hydraulic resistances of each microfluidic channel in the model ($ch_\mathrm{s}$, $ch_\mathrm{gi}$, $ch_\mathrm{go}$, and $ch_\mathrm{p}$, respectively), while $Q_\mathrm{s}$ and $Q_\mathrm{g}$ denote the volumetric flow rates of the applied average velocities from the supply inlet and gating inlet, respectively.

The dominance of the laminar flow regime in microfluidic systems allows us to apply Hagen-Poiseuille's law to our system \cite{mikaelian2020modeling}. This fundamental law states that the volumetric flow rate ($Q$) of the fluid in a microfluidic channel is directly proportional to the pressure difference ($\Delta P$) between the two ends of the channel, the fourth power of the channel's effective radius ($r$), and inversely proportional to the viscosity ($\mu$) of the fluid and the length ($l$) of the channel \cite{oh2012design}\cite{zhang2017logic} \cite{ajdari2004steady}:
\begin{equation}\label{Q}
Q = \frac{\pi  \Delta P  r^4}{8  \mu  l} .
\end{equation}
In rectangular channels, specifically in our model, the effective radius is equal to half of the hydraulic diameter and width of the channel ($w_\mathrm{ch}$), i.e., $r = \frac{D_h}{2} =\frac{w_\mathrm{ch}}{2}$ \cite{oh2012design}. With this consideration, \eqref{Q} can be rewritten as follows:
\begin{equation}\label{eq:Q}
Q = \frac{\pi   \Delta P  w_\mathrm{ch}^4}{128  \mu  l} = \frac{\Delta P}{R}.
\end{equation}
The above-mentioned equation is analogous to Ohm's law equation utilized in electrical circuits \cite{zhang2017logic}. In this analogy, the pressure difference between two points in the microfluidic channel, denoted as $\Delta P$, corresponds to the voltage difference between two points in the circuit. Similarly, the volumetric flow rate of the fluid in the channel ($Q$) is analogous to the current in electrical circuits. Furthermore, the hydraulic resistance of the channel, denoted as $R$, can be likened to electrical resistors in a circuit. The hydraulic resistance of the channel is directly influenced by the viscosity of the fluid flowing through it ($\mu$) and the length of the channel ($l$). It is inversely related to the width of the channel ($w_\mathrm{ch}$). The relationship between these parameters can be described by the following equation \cite{oh2012design}:
\begin{equation}\label{eq:R}
R = \frac{128  \mu  l}{\pi  w_\mathrm{ch}^4}.
\end{equation}
By utilizing the aforementioned equation, we can calculate the hydraulic resistance in each of the interconnected microfluidic channels in our model:
\begin{align}\label{eq:R1}
R_\mathrm{s} &= \frac{128  \mu  l_\mathrm{s}}{\pi  w_\mathrm{ch}^4} ,\\ \nonumber
R_\mathrm{gi} &= \frac{128  \mu  l_\mathrm{gi}}{\pi  w_\mathrm{ch}^4} ,\\ \nonumber
R_\mathrm{go} &= \frac{128  \mu  l_\mathrm{go}}{\pi  w_\mathrm{ch}^4} ,\\ \nonumber
R_\mathrm{p} &= \frac{128  \mu  l_\mathrm{p}}{\pi  w_\mathrm{ch}^4}.
\end{align}
There exists a direct relationship between the volumetric flow rate ($Q$) of fluid within a microfluidic channel, the average flow velocity ($u$), and the cross-sectional area of the channel ($A$) \cite{oh2012design}. This relationship can be expressed as
\begin{equation}\label{eq:Q2}
Q = A \times u .
\end{equation}
To determine the average flow velocity ($u_\mathrm{m}$) in channel $ch_\mathrm{p}$ of the system, the electrical analogy model needs to be solved in two distinct states: ON and OFF mode. This approach allows us to find the electrical current across resistance $R_\mathrm{p}$ in the circuit-equivalent model, which corresponds to the volumetric flow rate of the fluid ($Q_\mathrm{p}$) in channel $ch_\mathrm{p}$.

\subsubsection{\textbf{Gating ON Mode}}

In this mode, as schematically illustrated in Fig. \ref{fig: Hydrodynamic electrical equivalence} (a) and (c), the average velocity in y-direction  $\hat{u}_\mathrm{y} =u_\mathrm{g}$ is applied to the gating inlet, while the average velocity in x direction $\hat{u}_\mathrm{x} = u_\mathrm{s}$ is applied to the supply inlet. Consequently, considering the cross-sectional area of all channels as $A_\mathrm{ch}$ and the relative ratio between applied flow velocities from inlets as $r_\mathrm{u} = \frac{u_\mathrm{g}}{u_\mathrm{s}}$, the volumetric flow rates $Q_\mathrm{s}$ and $Q_\mathrm{g}$ in the microfluidic channels $ch_\mathrm{s}$ and $ch_\mathrm{gi}$ can be determined using the following equations:
\begin{equation}\label{eq:Q11}
Q_\mathrm{s} = A_\mathrm{ch} \times u_\mathrm{s},
\end{equation}
\begin{equation}\label{eq:Q22}
Q_\mathrm{g} = A_\mathrm{ch} \times r_\mathrm{u} \times u_\mathrm{s}.
\end{equation}
By solving the electrical circuit-equivalent model for the microfluidic system, we can derive the following equation:
\begin{equation}\label{eq:Q4}
Q_\mathrm{p} = A_\mathrm{ch} \times u_\mathrm{m}^\mathrm{ON}= \frac{R_\mathrm{go}  (Q_\mathrm{s} + Q_\mathrm{g})}{R_\mathrm{go} + R_\mathrm{p}}.
\end{equation}
By substituting \eqref{eq:Q11} and \eqref{eq:Q22} into \eqref{eq:Q4}, we can derive the following equation to determine the average velocity ($u_\mathrm{m}$) in channel $ch_\mathrm{p}$  during the system's ON mode:
\begin{equation}\label{eq:VM_ON}
u_\mathrm{m}^\mathrm{ON} = \frac{{R_\mathrm{go}  u_\mathrm{s}  (r_\mathrm{u} + 1)}}{{R_\mathrm{go} + R_\mathrm{p}}}.
\end{equation}

\subsubsection{\textbf{Gating OFF Mode}}

In this mode, which lasts for a duration of $T_\mathrm{g}$, $u_\mathrm{s}$ is the only applied flow velocity in the system. To solve the electric circuit-equivalent model in this mode (schematically illustrated in Fig. \ref{fig: Hydrodynamic electrical equivalence} (b)), we first define the equivalent resistance of the two parallel resistors $R_\mathrm{gi}$ and $R_\mathrm{go}$ as follows:
\begin{equation}\label{eq:REQ}
R_\mathrm{g} = \frac{{R_\mathrm{gi}  R_\mathrm{go}}}{{R_\mathrm{gi} + R_\mathrm{go}}}.
\end{equation}
Calculating the electrical current division between $R_\mathrm{g}$ and $R_\mathrm{p}$ resistors, and solving the circuit, we obtain the following equation:
\begin{equation}\label{eq:Q44444}
Q_\mathrm{p} = A_\mathrm{ch} \times u_\mathrm{m}^\mathrm{OFF}= \frac{{R_\mathrm{g}  Q_\mathrm{s}}}{{R_\mathrm{g} + R_\mathrm{p}}}.
\end{equation}
By substituting \eqref{eq:Q11} in \eqref{eq:Q44444}, we can calculate the average velocity ($u_\mathrm{m}^\mathrm{OFF}$) in channel $ch_\mathrm{p}$  during the system's OFF mode:
\begin{equation}\label{eq:VM_OFF_2}
u_\mathrm{m}^\mathrm{OFF} = \frac{{R_\mathrm{g}  u_\mathrm{s}}}{{R_\mathrm{g} + R_\mathrm{p}}}.
\end{equation}
\subsubsection{\textbf{Overall Average Velocity}}

Using the expressions for $u_\mathrm{m}^\mathrm{ON}$ and $u_\mathrm{m}^\mathrm{OFF}$, we can calculate the overall average velocity ($u_\mathrm{m}$) in channel $ch_\mathrm{p}$ . Within a period of $t$, the average velocity is $u_\mathrm{m}^\mathrm{OFF}$ in channel $ch_\mathrm{p}$ for a duration of $T_\mathrm{g}$, and $u_\mathrm{m}^\mathrm{ON}$ for a duration of $(t - T_\mathrm{g})$.Therefore, the overall average flow velocity $u_\mathrm{m}$ in channel $ch_\mathrm{p}$ can be determined as follows
\begin{equation}\label{eq:VM_VMONOFF}
u_\mathrm{m} = \frac{{(t - T_\mathrm{g})  u_\mathrm{m}^\mathrm{ON} + T_\mathrm{g}  u_\mathrm{m}^\mathrm{OFF}}}{{t}}.
\end{equation}
However, since the system is predominantly in the ON mode for generating concentration pulses ($ T_\mathrm{g} \ll t $), the terms $T_\mathrm{g} \times  u_\mathrm{m}^\mathrm{OFF}$ can be neglected in the above equation, and the overall flow velocity can be approximated as $u_\mathrm{m} \approx u_\mathrm{m}^\mathrm{ON}$. Considering $l_\mathrm{s}=l_\mathrm{gi}=l_\mathrm{go}$ in this system, we have:
\begin{equation}\label{eq:R3_R4}
R_\mathrm{go}+R_\mathrm{p}= \frac{128  \mu  l_\mathrm{ch}}{\pi  w_\mathrm{ch}^4},
\end{equation}
where $l_\mathrm{ch}$ represents the microfluidic channel length along the x-axis. By incorporating the above relation into \eqref{eq:VM_ON}, the overall $u_\mathrm{m}$ in this system can be expressed as follows:
\begin{equation}\label{eq:VM_FINALL_2}
u_\mathrm{m} = \frac{ l_\mathrm{go} u_\mathrm{s}  (r_\mathrm{u} + 1)}{l_\mathrm{ch}}.
\end{equation}

\subsubsection{\textbf{Approximate Analytical Expression for $W_\mathrm{g}$}}

By substituting \eqref{eq:VM_FINALL_2} into \eqref{wp1_tg__vm}, We obtain the following expression for $W_\mathrm{g}$ in our proposed model:
\begin{equation}\label{wp1_analytical}
 W_\mathrm{g}= \frac{ k_\mathrm{g}  T_\mathrm{g}  l_\mathrm{go} u_\mathrm{s}  (r_\mathrm{u} + 1)}{l_\mathrm{ch}}.
\end{equation}
\subsection{Pulse Propagation Compartment}
Referring to the schematic illustrated in Fig. \ref{fig:strategy}, the pulse propagation compartment receives inputs from the output of the pulse generation compartment, which are $A_\mathrm{g}$ and $W_\mathrm{g}$. The pulse propagation compartment processes these inputs and produces outputs for the entire system, which are $A_\mathrm{p}$ (pulse amplitude at $x_\mathrm{s}$), $W_\mathrm{p}$ (pulse width at $x_\mathrm{s}$) and $T_\mathrm{d}$ (pulse delay).
\begin{figure*}
     \centering
     \begin{subfigure}[b]{0.32\textwidth}
         \centering
          \includegraphics[width=\textwidth]{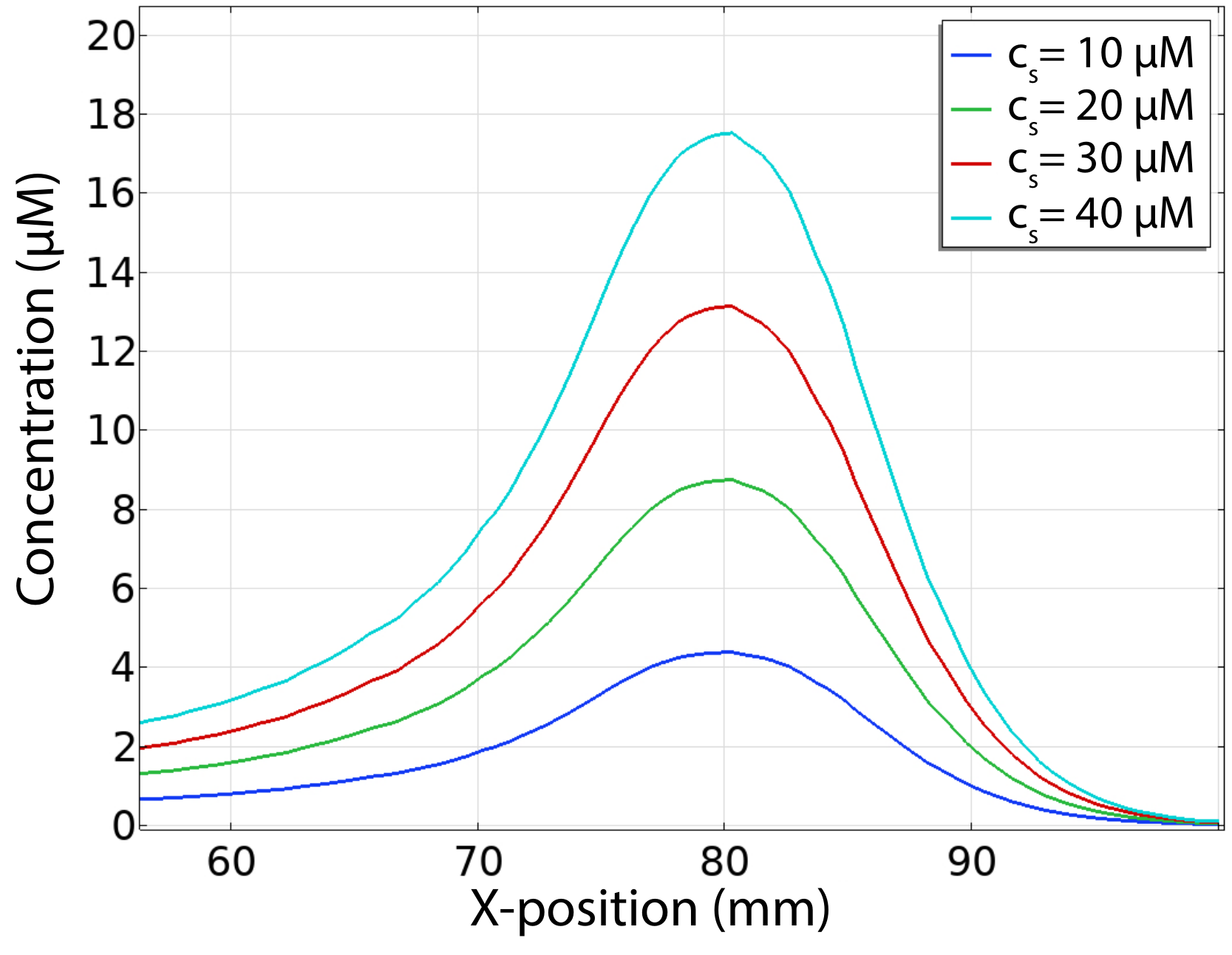}
         \caption{}
         \label{}
    \end{subfigure}
    \hfill
         \begin{subfigure}[b]{0.32\textwidth}
         \centering
          \includegraphics[width=\textwidth]{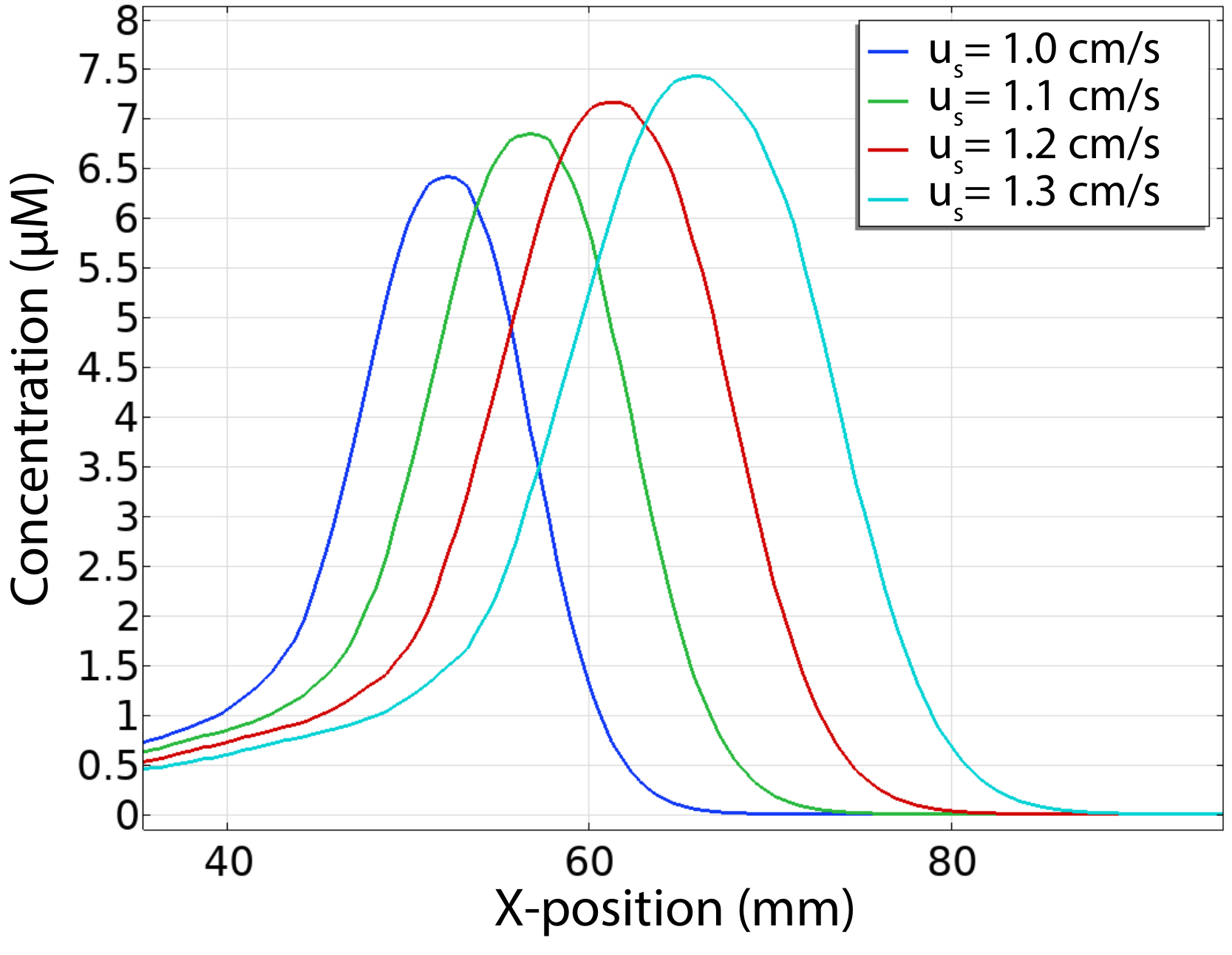}
         \caption{}
         \label{}
    \end{subfigure}
    \hfill
         \begin{subfigure}[b]{0.32\textwidth}
         \centering
          \includegraphics[width=\textwidth]{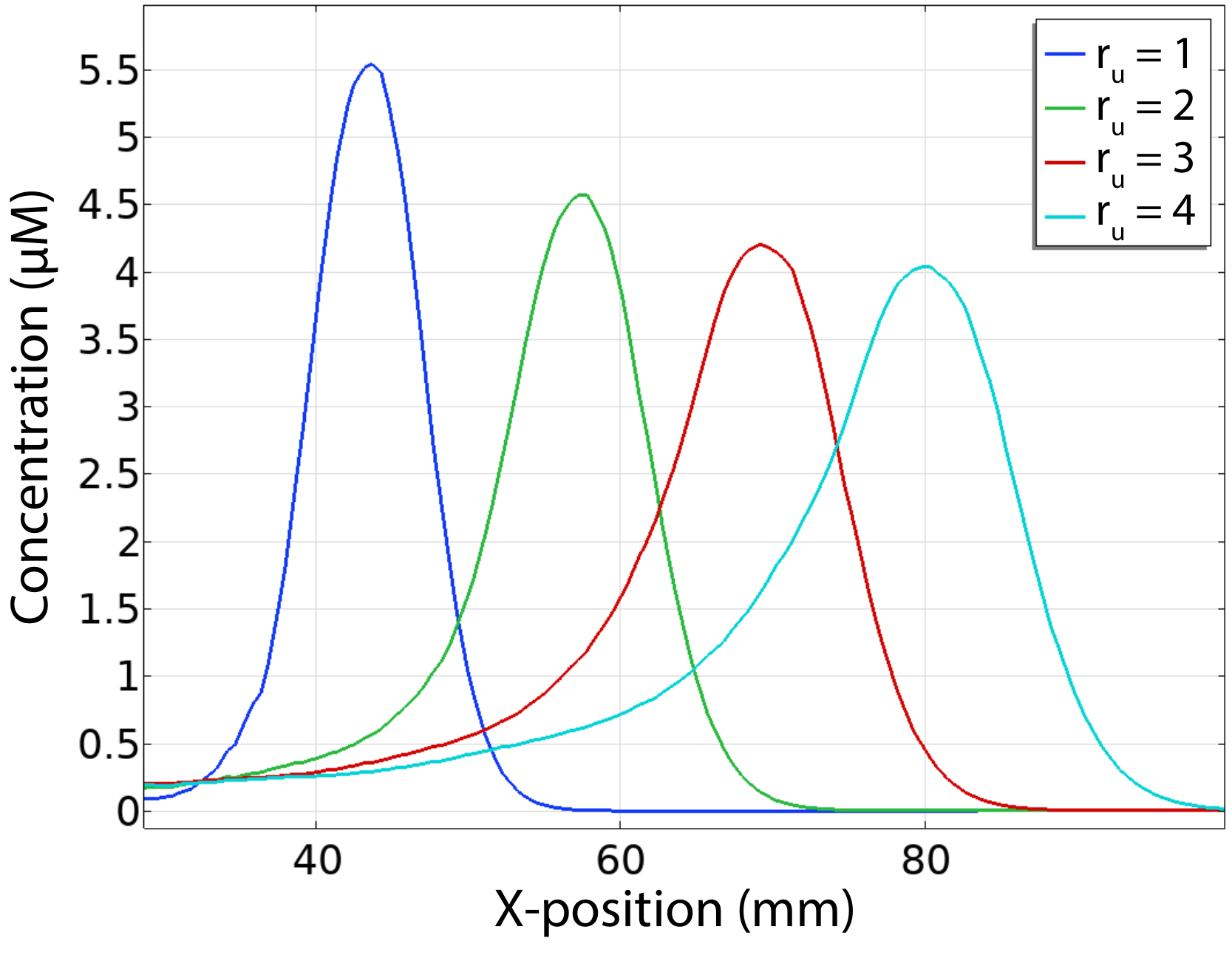}
         \caption{}
         \label{}
    \end{subfigure}

         \begin{subfigure}[b]{0.31\textwidth}
         \centering
          \includegraphics[width=\textwidth]{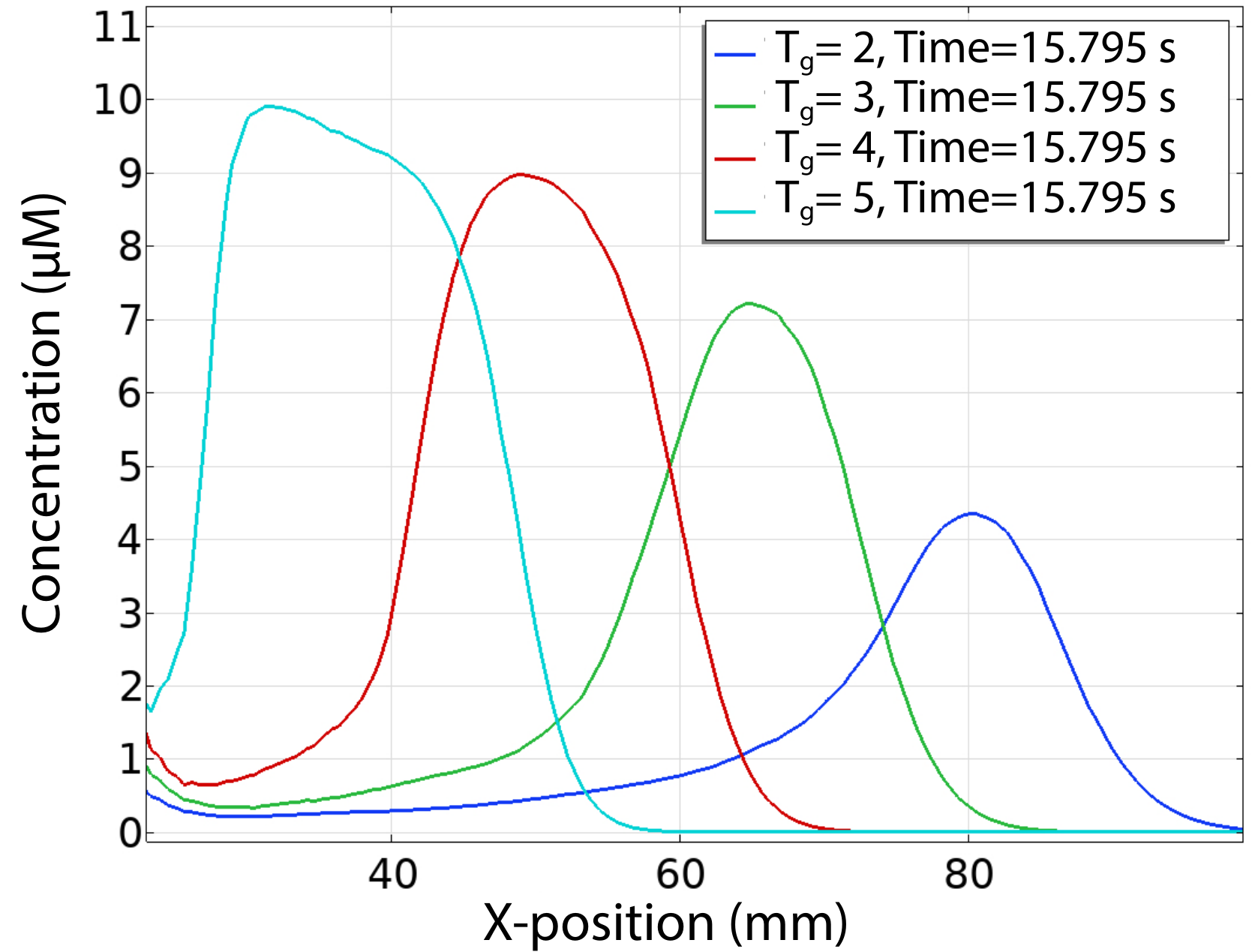}
         \caption{}
         \label{}
    \end{subfigure}
    \hfill
         \begin{subfigure}[b]{0.31\textwidth}
         \centering
          \includegraphics[width=\textwidth]{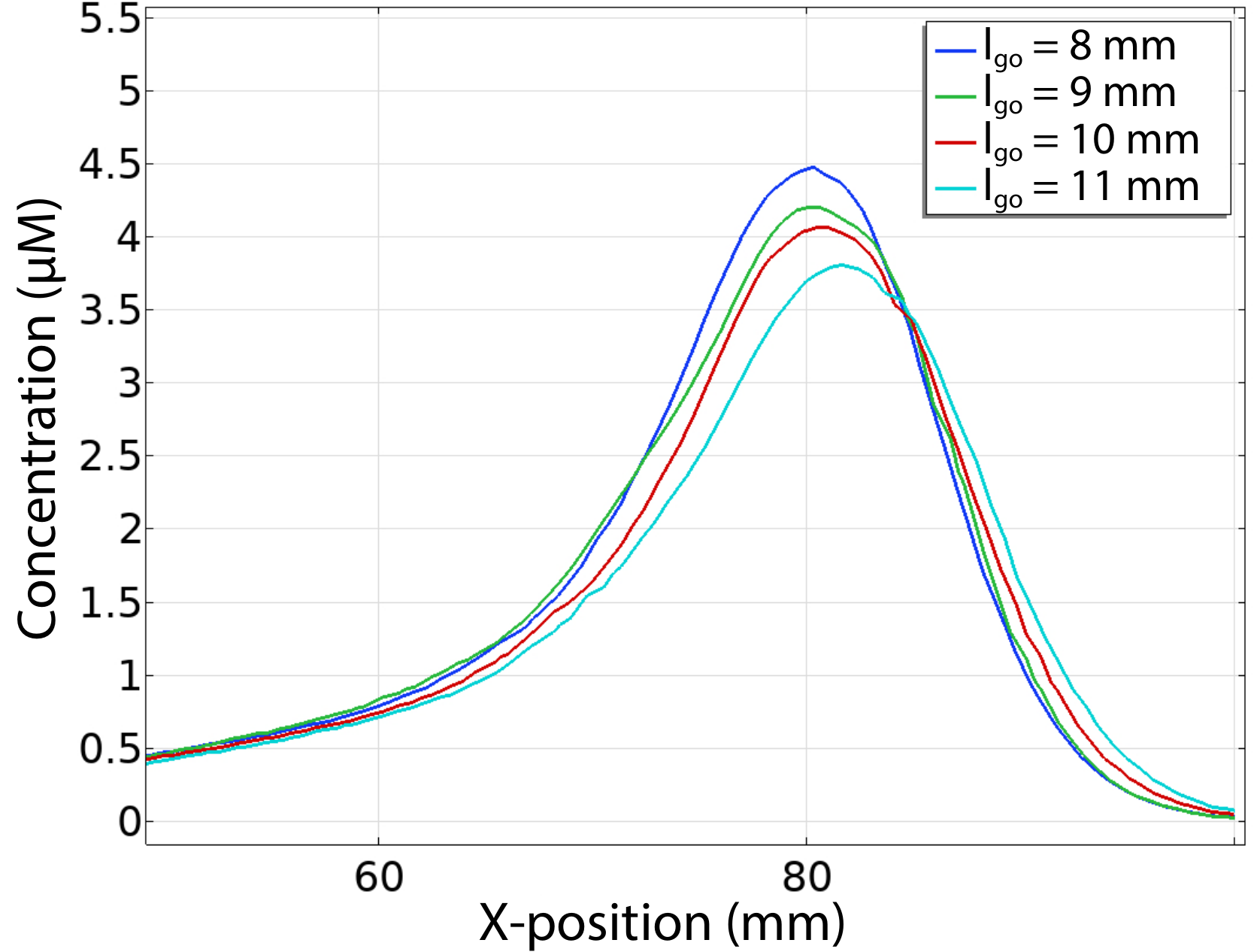}
         \caption{}
         \label{}
    \end{subfigure}
    \hfill
         \begin{subfigure}[b]{0.31\textwidth}
         \centering
          \includegraphics[width=\textwidth]{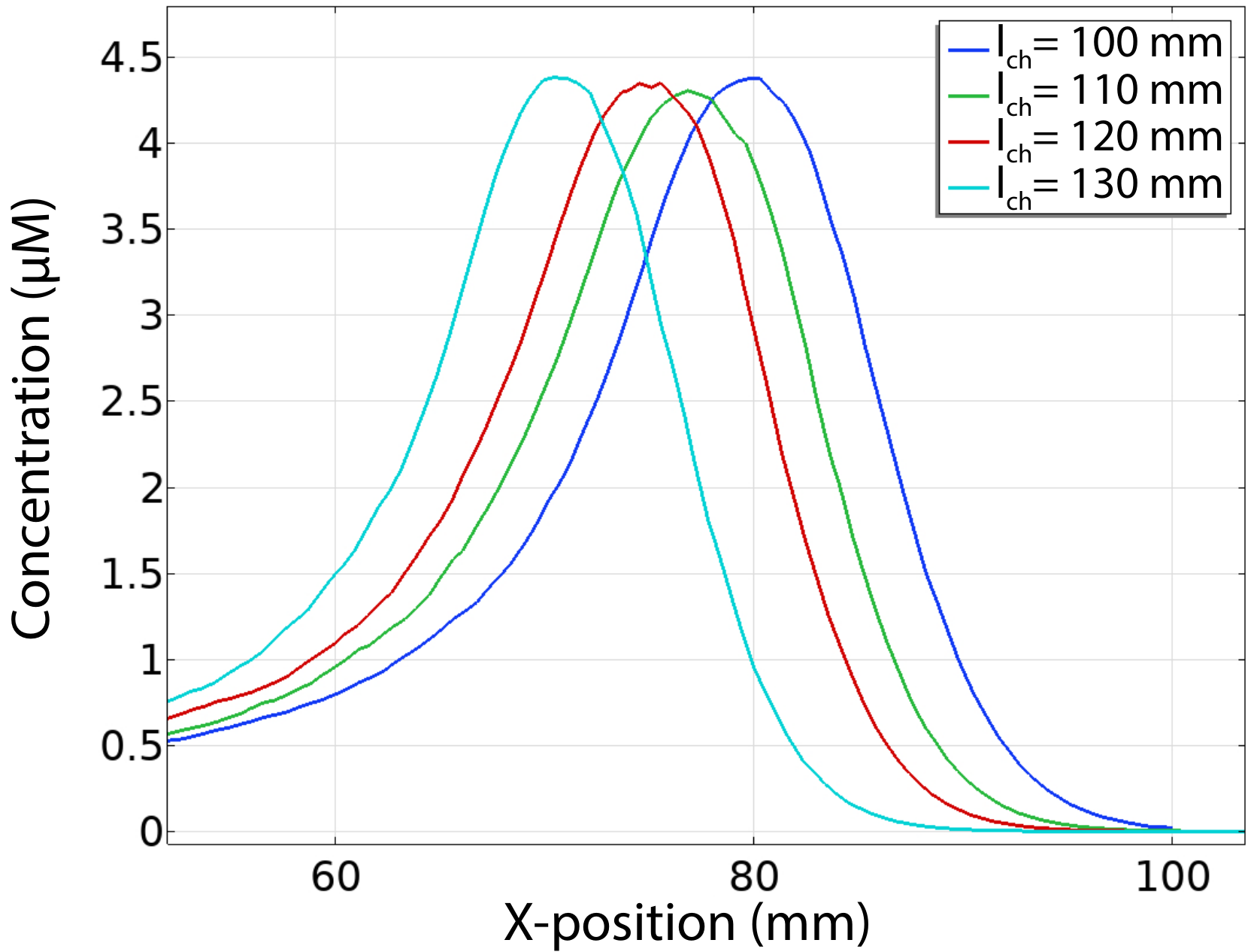}
         \caption{}
         \label{}
    \end{subfigure}
     \caption{Investigation of the impact of input parameters on transmitted Gaussian concentration pulse at the sampling point through parametric sweep simulation in COMSOL Multiphysics.}
         \label{fig: 6input effect diagram2}
    \end{figure*}

    \begin{figure*}
     \centering
     \begin{subfigure}[b]{0.32\textwidth}
         \centering
         \includegraphics[width=\textwidth]{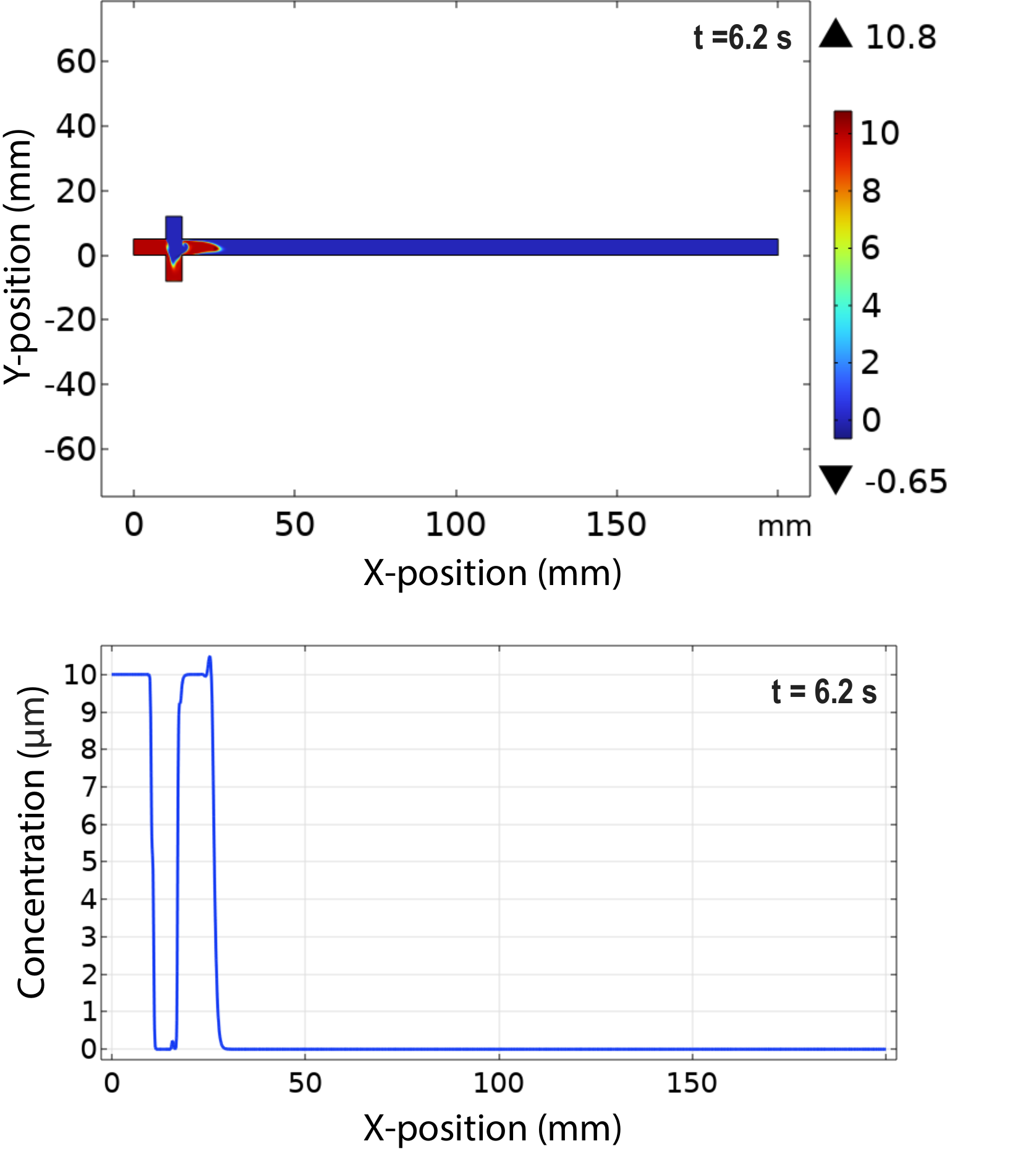}
         \caption{}
         \label{}
     \end{subfigure}
     \begin{subfigure}[b]{0.32\textwidth}
         \centering
         \includegraphics[width=\textwidth]{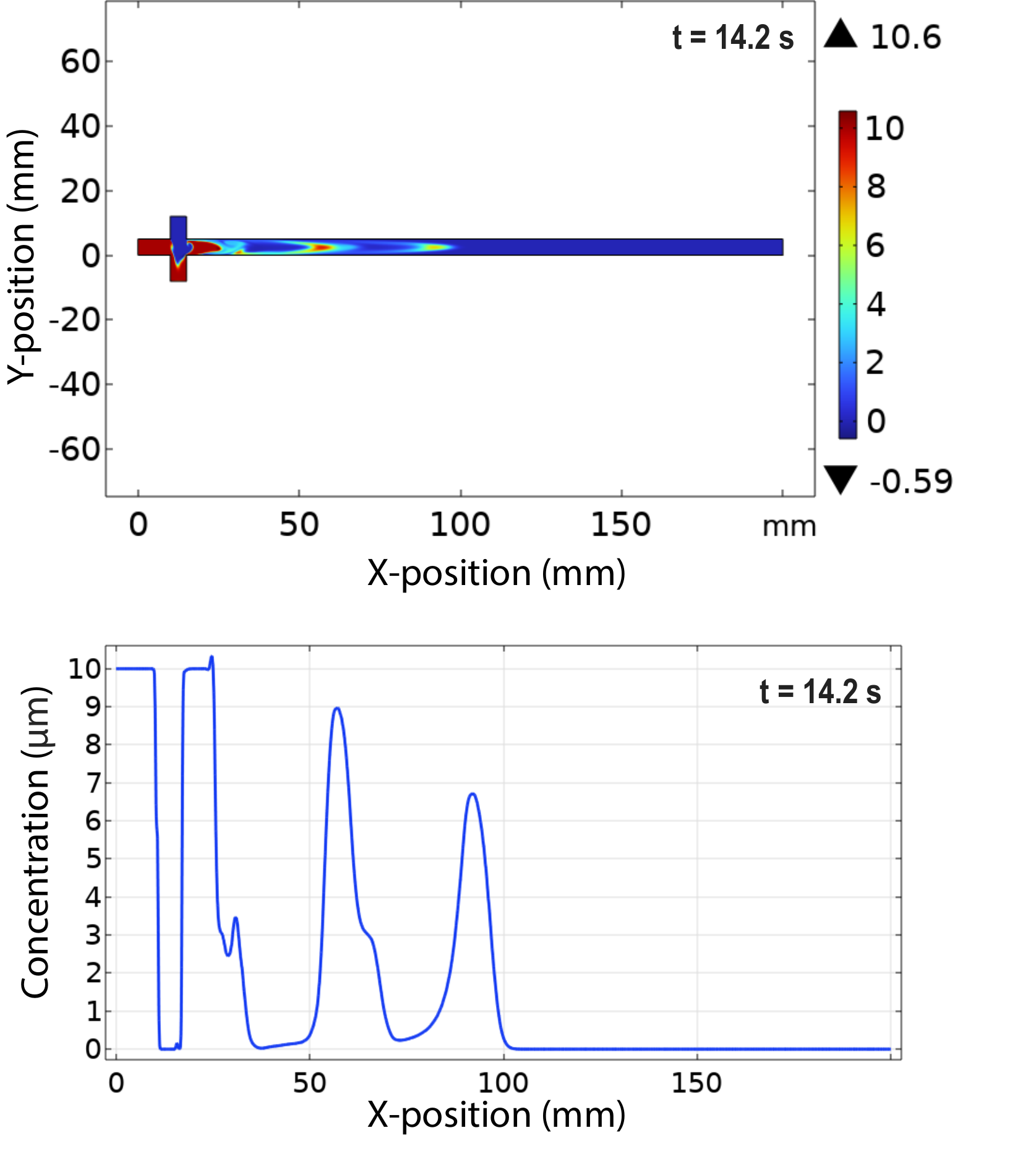}
         \caption{}
         \label{}
     \end{subfigure}
     \begin{subfigure}[b]{0.32\textwidth}
         \centering
         \includegraphics[width=\textwidth]{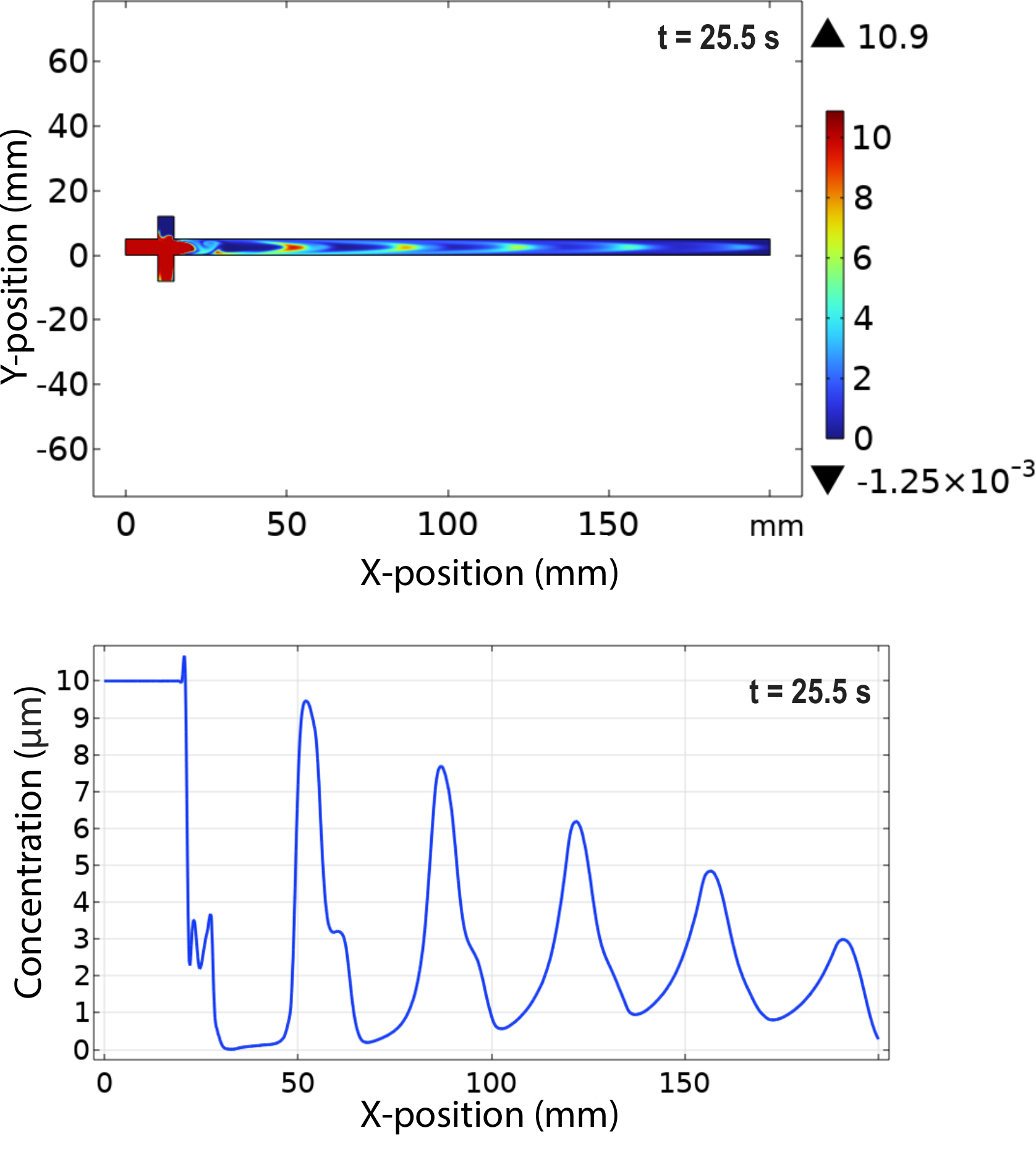}
         \caption{}
         \label{}
     \end{subfigure}
         \caption{An illustration of consecutive concentration pulses sampled at three different time instances: a) t = 6.20s, b) t = 14.20s, and c) t= 25.50s.}
         \label{fig:hydrodynamic3}
\end{figure*}
\subsubsection{\textbf{Approximate Analytical Expressions for $A_\mathrm{p}$ and $W_\mathrm{p}$}}
To establish a mathematical model governing the variables $A_\mathrm{p}$ and $W_\mathrm{p}$, we draw inspiration from physical phenomena observed in the propagation channel, as discussed in Section \ref{sec:exact model}. The concentration of information molecules at the pulse generation point, matching the initially applied concentration from the supply inlet ($c_\mathrm{s}$) according to \eqref{AP1}, propagates through the propagation channel. The pulse amplitude of the propagating pulse at the sampling point depends on the initial concentration at the pulse generation point, where an increase in the $c_\mathrm{s}$ leads to a higher $A_\mathrm{p}$. Moreover, the pulse amplitude of the propagating pulse ($A_\mathrm{p}$) is correlated with the sampling point ($x_\mathrm{s}$), and there will be a gradual decrease in the peak value of the pulse during propagation in the channel due to the gradual dilution of the molecules in the medium. Additionally, Higher flow velocity along the propagation channel contributes more to the convection phenomenon in the channel, leading to an increase in the number of propagating molecules through the propagation channel and consequently resulting in an increase in $A_\mathrm{p}$. Furthermore, since our derived expression for $W_\mathrm{g}$ is dependent on $u_\mathrm{m}$ in $ch_\mathrm{p}$ channel, there exists a direct relationship between $A_\mathrm{p}$ and $W_\mathrm{g}$, where an increase in $W_\mathrm{g}$ will lead to a higher value of $A_\mathrm{p}$.
As the concentration of the generated pulse is always equal to $c_\mathrm{s}$, the pulse width of the generated pulse plays a critical role in determining the number of molecules entering the propagation channel. As a result, there exists a linear relationship between the pulse width of the generated pulse and the pulsed width of the propagating pulse. Likewise, the pulse width of the propagating pulse is dependent on the sampling point, as it varies at each point along the channel due to the dispersion effect occurring in the channel. Based on arisen intuition from the physical phenomena observed in the model system, we can express the following expressions for the $A_\mathrm{p}$ and $W_\mathrm{p}$:
\begin{equation}\label{ap2_2}
A_\mathrm{p} = \frac{k_\mathrm{a} A_\mathrm{g} W_\mathrm{g} }{(x_\mathrm{s}-x_\mathrm{g})},
\end{equation}
\begin{equation}\label{wp2_2}
W_\mathrm{p} = k_\mathrm{w}(x_\mathrm{s}-x_\mathrm{g}) W_\mathrm{g},
\end{equation}
where $k_\mathrm{a}$ and $k_\mathrm{w}$ are fitting parameters that will be determined at the end of this section (the optimization part).
\subsubsection{\textbf{Approximate Analytical Expression for $T_\mathrm{d}$}}
To establish a mathematical model governing the variable $T_\mathrm{d}$, we draw upon insights from the physical phenomena observed in the model system, as discussed in Section \ref{sec:exact model}. An increase in the flow velocity ($u_\mathrm{m}$) in the propagation channel ($ch_\mathrm{p}$), results in a rapid propagation of information molecules due to the convection effect. Hence, we can establish an inverse relationship between $u_\mathrm{m}$ and $T_\mathrm{d}$. Moreover, it is crucial to note that $T_\mathrm{d}$ also exhibits a relationship with the sampling point ($x_\mathrm{s}$) along the propagation channel. If the sampling point is chosen in close proximity to the pulse generation point, it will take less time for the pulse to reach the pulse sampling point, resulting in a reduced time delay. Therefore, the analytical expression for $T_\mathrm{d}$ in our proposed model is approximated as follows:
\begin{equation}\label{td}
T_\mathrm{d}=\frac{k_\mathrm{t} (x_\mathrm{s} - x_\mathrm{g})}{u_\mathrm{m}},
\end{equation}
where $k_\mathrm{t}$ is a fitting parameter that will be determined at the end of this section (the optimization part). 
\subsubsection{\textbf{Finite Element Simulation}}
In order to validate the accuracy of the developed analytical model in approximating the behavior of the sampled concentration pulse, comprehensive finite element simulations were performed using COMSOL Multiphysics. The results of this analysis are provided in Fig. \ref{fig: 6input effect diagram2}. By systematically varying input parameters and observing their effects on the concentration pulse at specific sampling points, we confirmed the validity of our derived analytical expressions.
\subsubsection{\textbf{Model Optimization}}
\begin{table}[]
    \centering
    \caption{The simulated 30 scenarios that are used for the optimization of the model. In each scenario, the value of only one parameter is changed from the default setting to the value listed in the table.}
    \scalebox{0.8}{
    \begin{tabular}{|c|c|c|c|c|c|}
    \hline
        $c_\mathrm{s}$&$1$mM&$10$mM&$20$mM&$30$mM&$40$mM\\          
        \hline
        $u_\mathrm{s}$ & $10$mm/s & $11$mm/s & $12$mm/s & $13$mm/s & $15$mm/s\\
        \hline
        $r_\mathrm{u}$& $1$ & $2$ & $3$ & $4$ & $10$\\
        \hline
        $T_\mathrm{g}$& $1$s & $2$s & $3$s & $4$s & $5$s\\
        \hline
                 $l_\mathrm{ch}$ & $100$mm  & $110$mm  & $120$mm  & $130$mm  & $200$mm\\
         \hline
        $l_\mathrm{go}$& $12.5$mm & $13.5$mm &$14.5$mm & $15.5$mm & $25$mm\\
         \hline
    \end{tabular}}
    \label{scenarios}
\end{table}

\begin{table}[!b]
    \centering
    \caption{Optimal values of fitting parameters obtained through genetic algorithm.}
    \scalebox{0.9}{
    \begin{tabular}{|c|c|c|c|}
    \hline
        $k_\mathrm{g}=14.6$&$k_\mathrm{a}=20.16$&$k_\mathrm{w}=0.7$&$k_\mathrm{t}=0.683$ \\          
        \hline
    \end{tabular}}
    \label{constant_parameters}
\end{table}
To optimize and obtain the optimal values for the fitting parameters ($k_\mathrm{g}$, $k_\mathrm{a}$, $k_\mathrm{w}$ and $k_\mathrm{t}$) in our analytical model, we introduce an error function ($E$) that facilitates the optimization process. This function calculates the mean error for 30 different scenarios by comparing the results of the analytical and simulation models in terms of the three pulse parameters (pulse amplitude, width, and delay). In each of these scenarios, only one input parameter is changed from its default setting at a time. The details of the selected scenarios are presented in Table \ref{scenarios}. In each of these scenarios, we systematically varied one parameter from its default setting to five different values, ensuring a gradual increase up to the highest amount that our simulation model is able to generate concentration pulses effectively. During this process, we kept all other parameters constant. Our main objective in selecting these scenarios is to determine the ideal range of parameter values that guarantee the model's performance aligns closely with the simulation results. These scenarios play a pivotal role in defining the optimal limitations of our model. However, as we venture beyond specific parameter values, the model's accuracy diminishes and the model does not accurately capture the scenarios beyond the selected range of parameters. The expression for the error function ($E$) is represented as follows:
\begin{align}\label{E}
E &= \frac{1}{N} \sum_{i=1}^{N} \left(|A_\mathrm{p_{\mathrm{sim},i}} - A_\mathrm{p_{\mathrm{ana},i}}| + |W_\mathrm{p_{\mathrm{sim},i}} - W_\mathrm{p_{\mathrm{ana},i}}| \right.\\ \nonumber
&\qquad \left. + |T_\mathrm{d_{\mathrm{sim},i}} - T_\mathrm{d_{\mathrm{ana},i}}|\right),
\end{align}
where $N$ denotes the total number of scenarios under consideration, and $A_\mathrm{p_{\mathrm{sim},i}}$, $W_\mathrm{p_{\mathrm{sim},i}}$, and $T_\mathrm{d_{\mathrm{sim},i}}$ represent the pulse amplitude, width, and delay, respectively, obtained through finite element simulations for the $i^\mathrm{th}$ scenario, while, $A_\mathrm{p_{\mathrm{ana},i}}$, $W_\mathrm{p_{\mathrm{ana},i}}$, and $T_\mathrm{d_{\mathrm{ana},i}}$ represent the corresponding values obtained from the analytical model for the $i^\mathrm{th}$ scenario. Accordingly, we have obtained the optimal values of the fitting parameters that minimize the error function by using a genetic algorithm optimization in MATLAB. These values are presented in Table \ref{constant_parameters}.

\section{Successive Pulse Transmission Model}
\label{sec:successive pulse Transmitter}
\begin{figure}[t]
\centering
\includegraphics[width=0.45\textwidth]{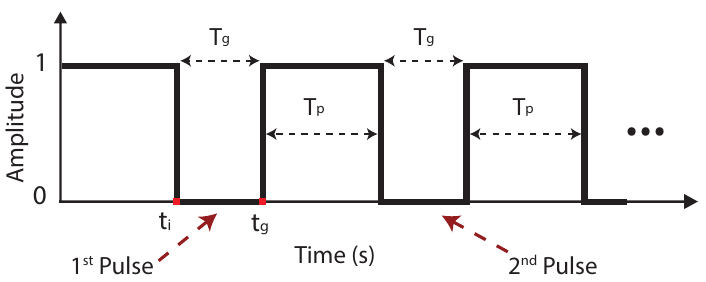}
\caption{Applied square pulse on the gating inlet in successive-pulse generation model.}
    \label{fig:multi pulse square}
\end{figure}

To observe the ISI arising from significant memory effects inherent in the diffusion-based MC channel, leading to time-varying and stochastic offset to the sampled signals, we conducted simulations for consecutive generation and transmission of concentration pulses using the hydrodynamic gating technique. This involved applying consecutive square pulses to the gating inlet, each with a duration of $T_\mathrm{g}$ seconds and a time interval of $T_\mathrm{p}$ between each pulse, as depicted in Fig. \ref{fig:multi pulse square}. In Fig. \ref{fig:hydrodynamic3}, we present the 1D and 2D time-evolving concentration profiles of five consecutive concentration pulses generated with the same gating durations, and even pulse generation intervals. The effect of dispersion on the width and peak amplitude of the individual concentration pulses can be observed.
Although the pulse peaks are easily distinguishable for this setup with a particular channel length, it becomes evident that the interference of consecutive pulses increases as they approach the end of the microfluidic channel. These results highlight the importance of generating short concentration pulses to avoid or minimize the ISI.

\begin{figure*}[t!]
     \centering
     \begin{subfigure}[b]{0.32\textwidth}
         \centering
          \includegraphics[width=\textwidth]{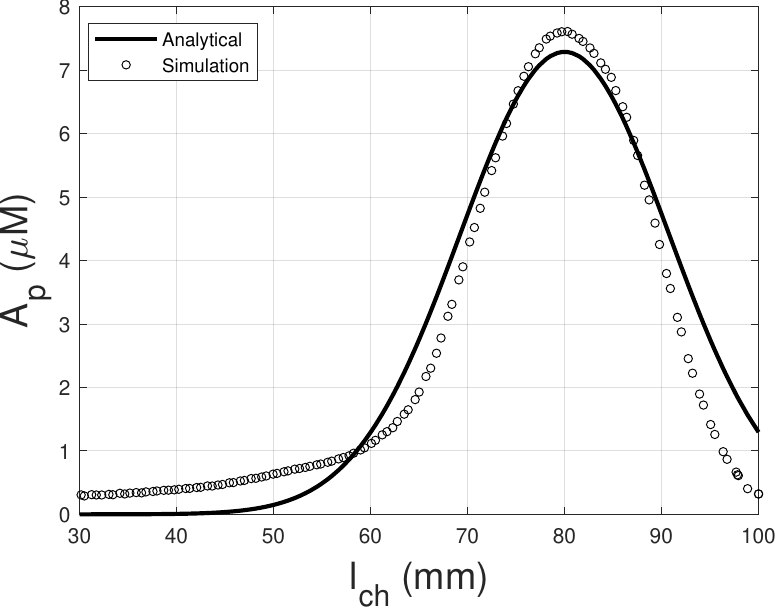}
         \caption{}
         \label{}
    \end{subfigure}
    \hfill
     \begin{subfigure}[b]{0.32\textwidth}
         \centering
          \includegraphics[width=\textwidth]{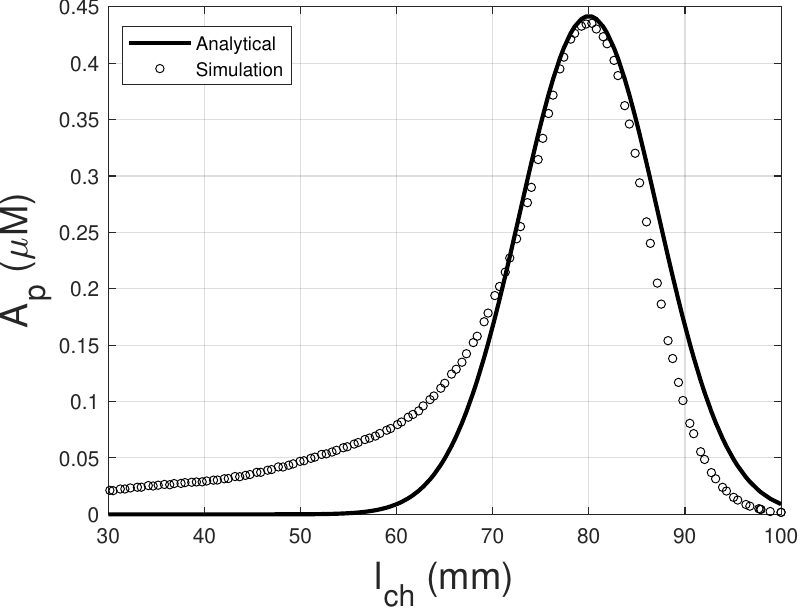}
         \caption{}
         \label{}
    \end{subfigure}
\hfill
       \begin{subfigure}[b]{0.32\textwidth}
         \centering
          \includegraphics[width=\textwidth]{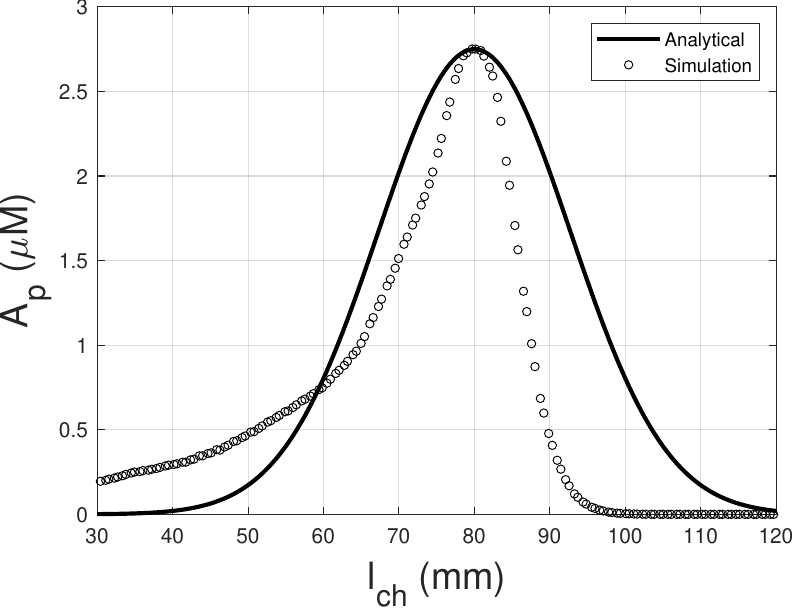}
         \caption{}
         \label{}
    \end{subfigure}

         \begin{subfigure}[b]{0.32\textwidth}
         \centering
          \includegraphics[width=\textwidth]{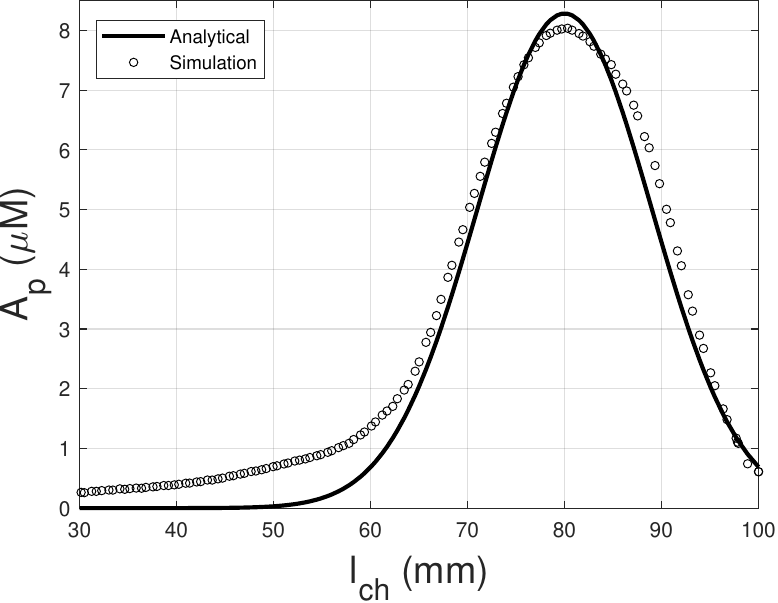}
         \caption{}
         \label{}
    \end{subfigure}
    \hfill
       \begin{subfigure}[b]{0.32\textwidth}
         \centering
          \includegraphics[width=\textwidth]{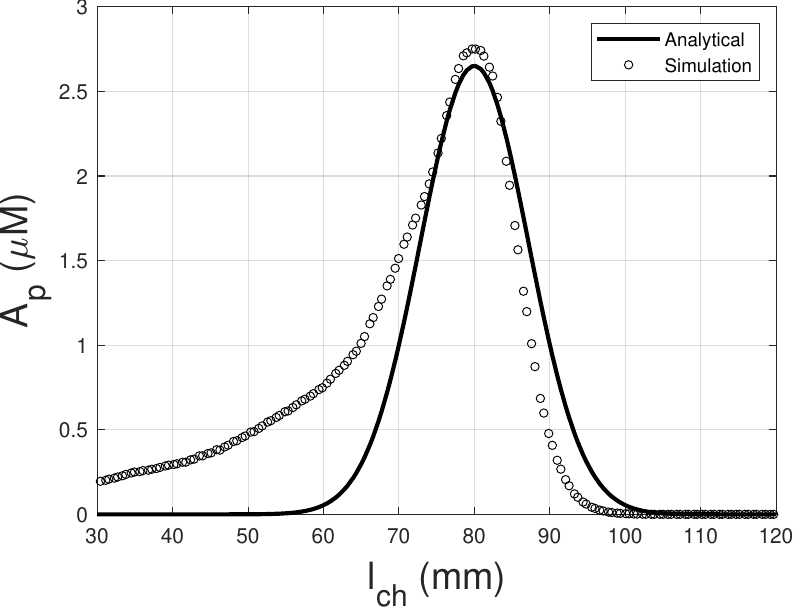}
         \caption{}
         \label{}
    \end{subfigure}
       \hfill 
         \begin{subfigure}[b]{0.32\textwidth}
         \centering
          \includegraphics[width=\textwidth]{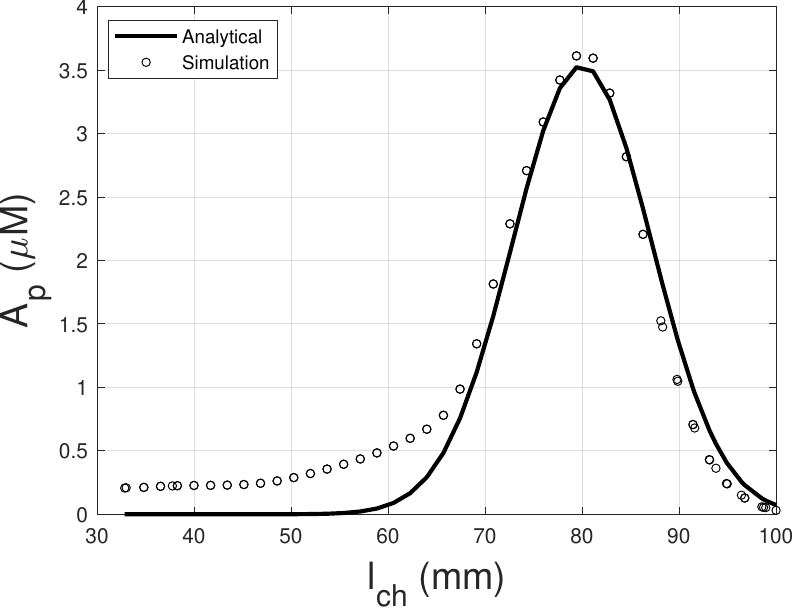}
         \caption{}
         \label{}
    \end{subfigure}
    \hspace{2mm}
             \caption{\textbf{Concentration profile of the sampled pulse at $(x_\mathrm{s},t_\mathrm{s})$ for varying system parameters.} In each analysis, only one parameter is changed from its default value given in Table \ref{table:inputs}. a) $c_\mathrm{s}= 10^{-1}\mu$M, b) $u_\mathrm{s} = 1.5 \times 10^{-2} $m/s, c) $r_\mathrm{u}= 10$, d) $T_\mathrm{g} = 5$s, e) $l_{\mathrm{ch}} = 200$mm, f) $l_{\mathrm{go}} = 25$mm. }
         \label{received_pulse}
\end{figure*}

To determine the mathematical model of the system in this mode, we utilized the equations derived from the single pulse transmission mode in the preceding section. We assumed that the transmission of the first pulse corresponds precisely to the analytical expression obtained for the single pulse transmission model. Subsequently, by neglecting the ISI effect, we obtained the pulse characteristics of the consecutive pulses based on the pulse characteristics of the first pulse. Each successive pulse is considered a time-shifted version of the first pulse, with a time shift of $T_\mathrm{p}$. Therefore, we can derive the shifted distance occurring within the time span of $T_\mathrm{p}$ as follows:

\begin{equation}\label{eq33}
d = \frac{ T_\mathrm{p}(x_\mathrm{s} - x_\mathrm{g})}{T_\mathrm{d}},
\end{equation}
\begin{equation}
x_n=x_\mathrm{s}-(n-1)d, \hspace{5mm}n={1,2,3,...},
\end{equation}
where $d$ represents the shifted distance between the consecutive pulses, $x_n$ denotes the peak point of each consecutive pulse, and $n$ represents the number of consecutive pulses generated, starting from $n = 1$, where the first pulse corresponds to $n = 1$, consistent with the single-pulse model discussed in the previous section. By utilizing the derived values, the concentration profile of each consecutive pulse can be derived from the below equation:
\begin{equation}
C_n= A_\mathrm{p}|_{x=x_\mathrm{n}} \exp \left(-\frac{(x-x_\mathrm{n})^2}{2(W_\mathrm{p}|_{x=x_\mathrm{n}})^2}\right),
\end{equation}
where $C_n$ is the concentration profile of each time-shifted consecutive pulse. Therefore, the analytical model of our system in successive pulse generation mode can be obtained through a summation of time-shifted signals as 
\begin{equation}
\label{C_successive}
\left.C(x)\right|_{x=x_\mathrm{s}} = \sum_{n=1}^{N} C_n,
\end{equation}
with $N$ representing the total number of transmitted pulses.

\section{Results and Discussions}\label{section_results}
In this section, we evaluate the accuracy of the proposed analytical model by comparing the results for various scenarios to those obtained in finite element simulations in COMSOL Multiphysics. The default values for the system parameters used in these analyses, were already provided in Table \ref{table:inputs} in Section \ref{sec:hydrodynamic gating model}.

\begin{figure*}[t!]
     \centering
     \centering
     \begin{subfigure}[b]{0.32\textwidth}
         \centering
         \includegraphics[width=\textwidth]{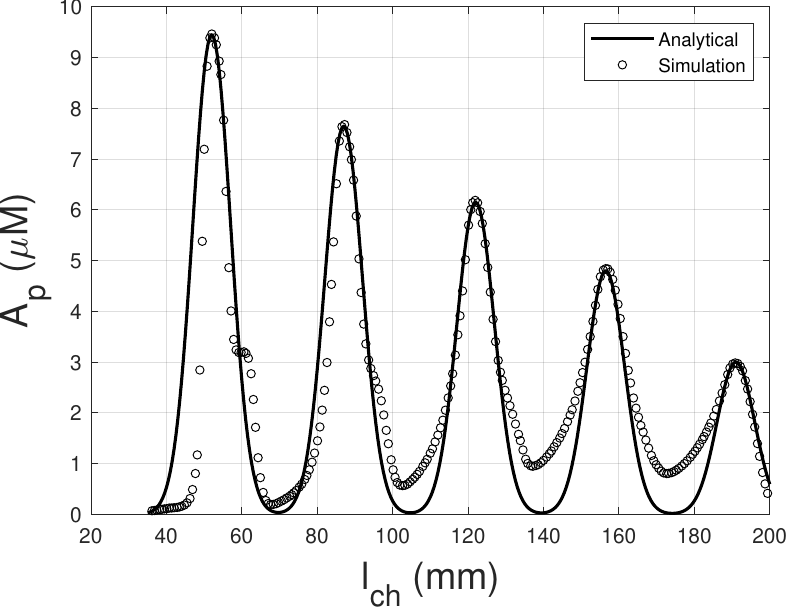}
         \caption{}
         \label{}
     \end{subfigure}
\hfill 
     \begin{subfigure}[b]{0.32\textwidth}
         \centering
         \includegraphics[width=\textwidth]{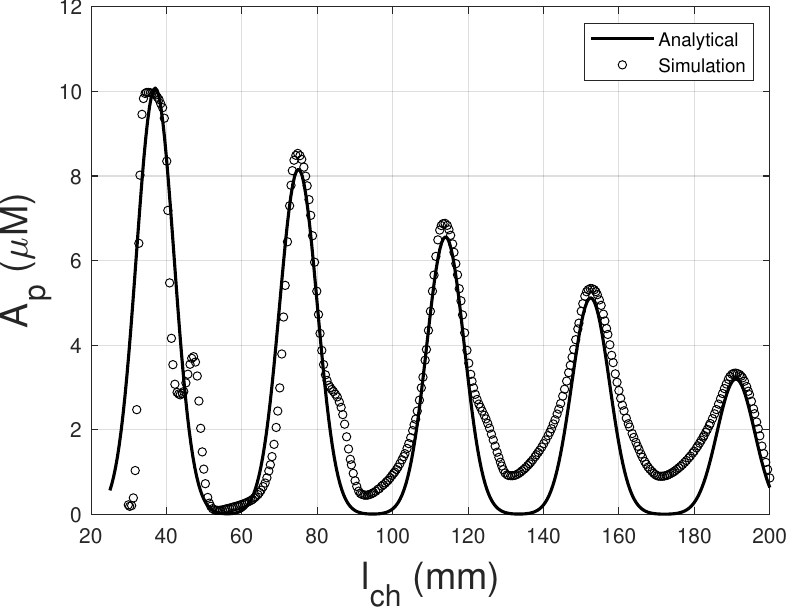}
         \caption{}
         \label{}
     \end{subfigure}
\hfill
     \centering
     \begin{subfigure}[b]{0.32\textwidth}
         \centering
         \includegraphics[width=\textwidth]{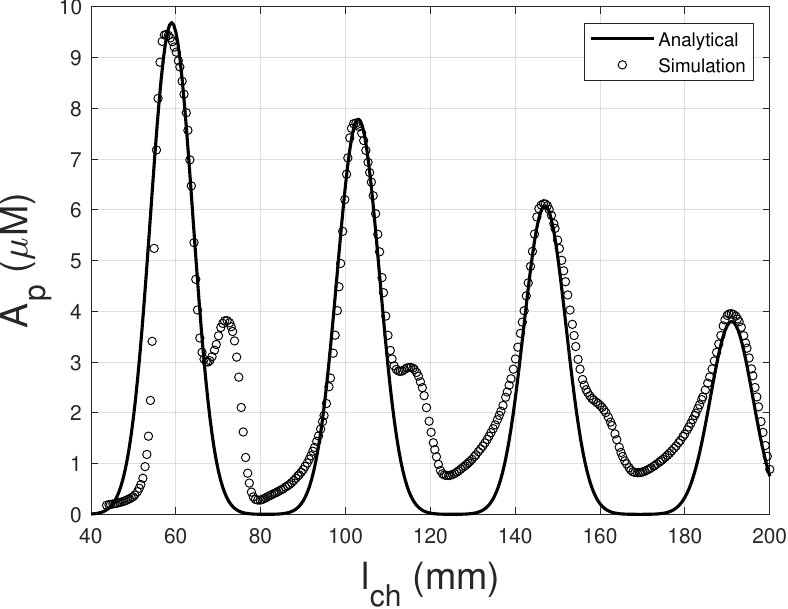}
         \caption{}
         \label{}
     \end{subfigure}
         \caption{\textbf{Concentration profile of successive pulses at $(x_\mathrm{s},t_\mathrm{s})$ for varying system parameters.} In each analysis, two parameters are changed from their default value given in Table \ref{table:inputs}. a) $u_\mathrm{s} = 1.3 \times 10^{-2} $m/s, $r_\mathrm{u}= 3$, b) $l_{\mathrm{go}} = 13.5$mm, $T_\mathrm{g} = 3$s c) $r_\mathrm{u}= 5$,, $u_\mathrm{s} = 1.5 \times 10^{-2} $m/s. }
         \label{successive_pulse}
     \end{figure*}

We conduct a comparison between the amplitude (concentration) of the sampled pulse at $(x_\mathrm{s},t_\mathrm{s})$ using \eqref{Gaussian1} under various scenarios, in each
of which we vary only one parameter from its default setting at a time and the corresponding numerical results obtained from experiments. The results, provided in Fig. \ref{received_pulse}, demonstrate an agreement between our analytical model and the numerical solution, thereby confirming the accuracy of the model. Additionally, the experiments are repeated for the successive pulse transmission model under various scenarios. Fig. \ref{successive_pulse} shows the results obtained using  \eqref{C_successive} and the finite element simulations. Clearly, the successive pulse model also exhibits a high level of agreement with the simulation results.

\begin{figure*}[t!]
     \centering
     \begin{subfigure}[b]{0.32\textwidth}
         \centering
         \includegraphics[width=\textwidth]{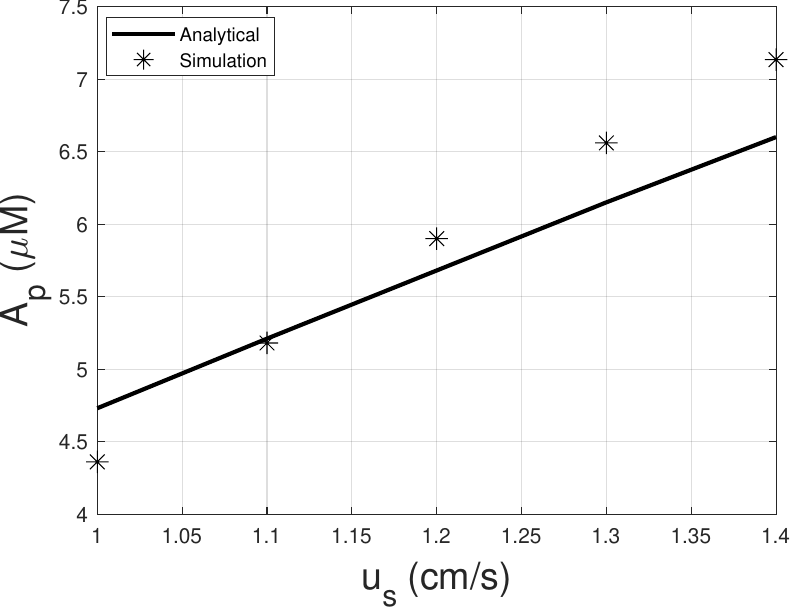}
         \caption{}
         \label{}
     \end{subfigure}
\hfill
          \begin{subfigure}[b]{0.32\textwidth}
         \centering
         \includegraphics[width=\textwidth]{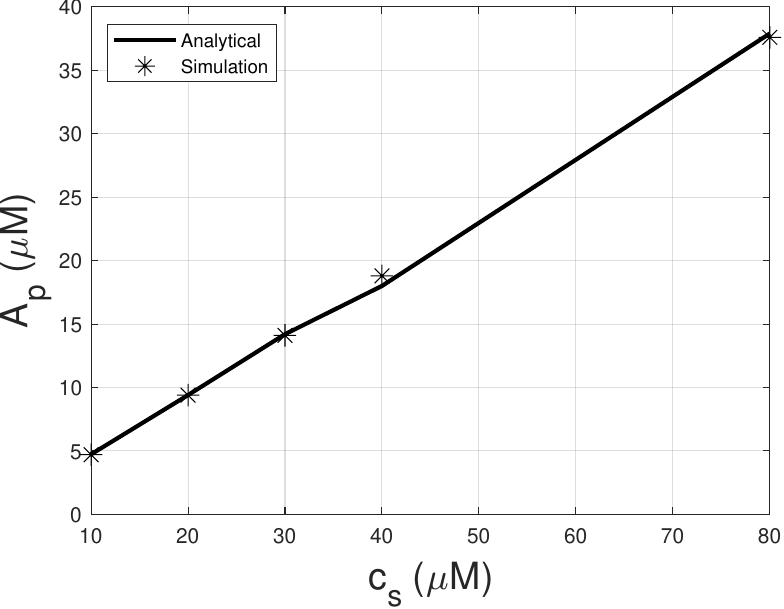}
         \caption{}
         \label{}
     \end{subfigure}
     \hfill 
               \begin{subfigure}[b]{0.32\textwidth}
         \centering
         \includegraphics[width=\textwidth]{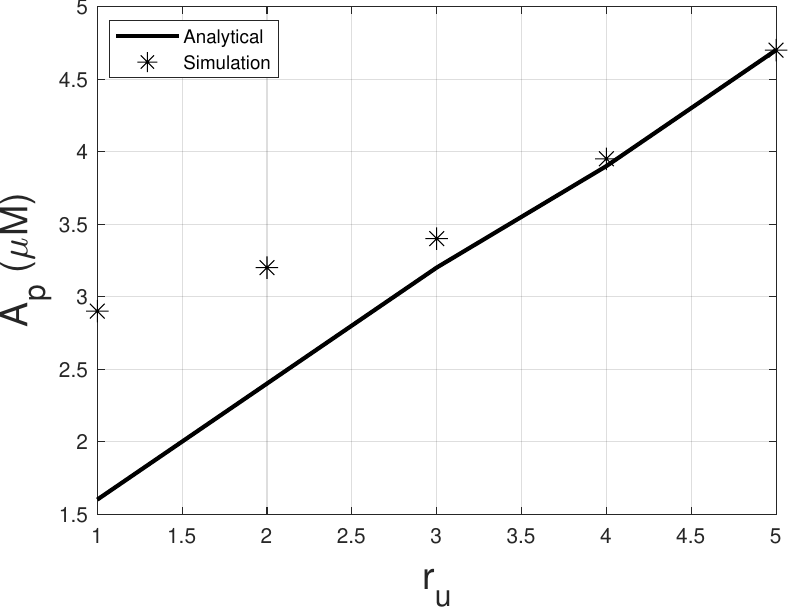}
         \caption{}
         \label{}
     \end{subfigure} 

               \begin{subfigure}[b]{0.32\textwidth}
         \centering
         \includegraphics[width=\textwidth]{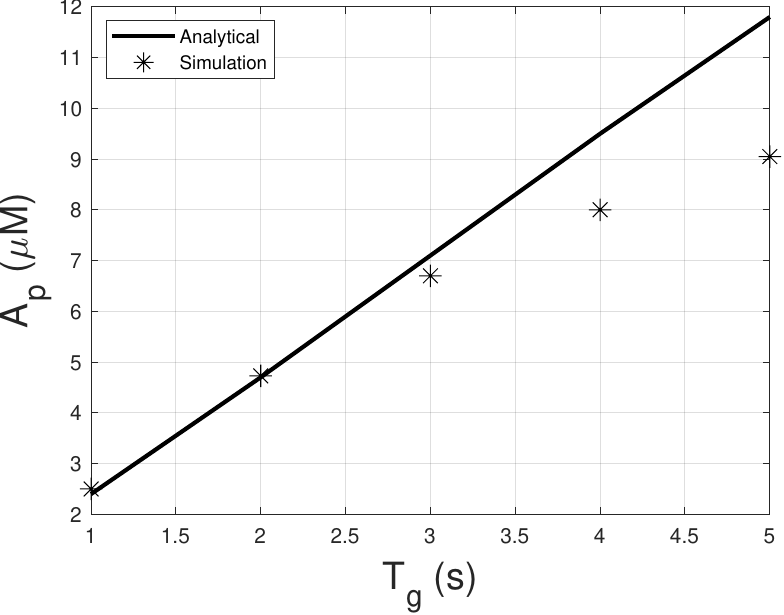}
         \caption{}
         \label{}
     \end{subfigure}
\hfill
               \begin{subfigure}[b]{0.32\textwidth}
         \centering
         \includegraphics[width=\textwidth]{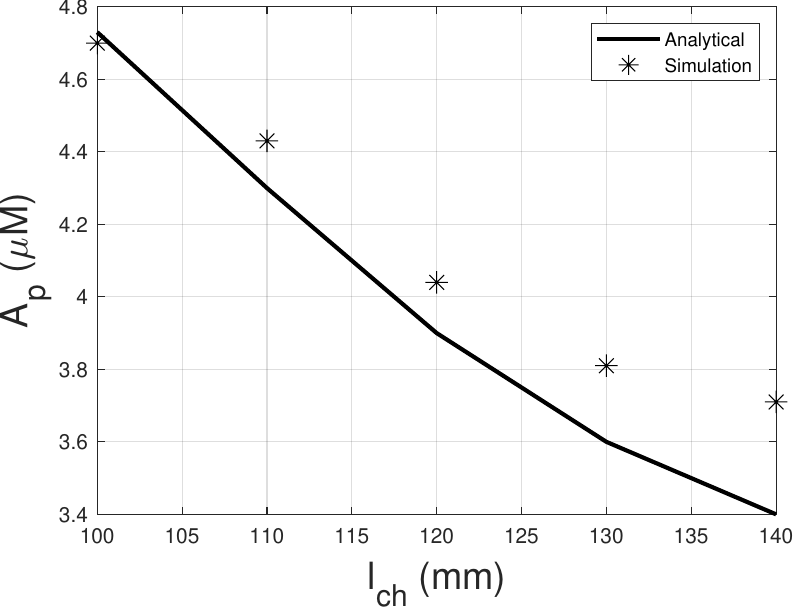}
         \caption{}
         \label{}
     \end{subfigure}
\hfill
          \begin{subfigure}[b]{0.32\textwidth}
         \centering
         \includegraphics[width=\textwidth]{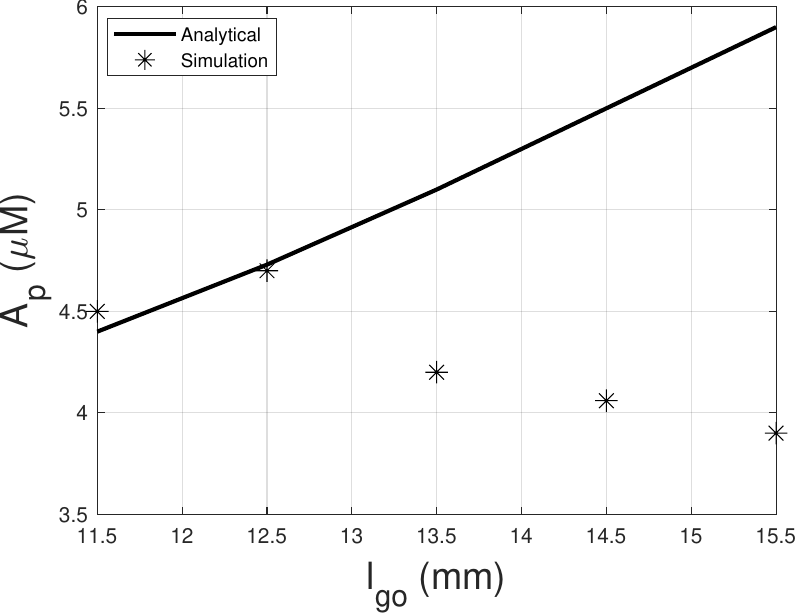}
         \caption{}
         \label{}
     \end{subfigure}
           \caption{\textbf{Pulse amplitude, $A_\mathrm{p}$, for varying system input parameters:} a) flow velocity, $u_\mathrm{s}$, b) input concentration, $c_\mathrm{s}$, c) ratio between velocity of supply flow and gating flow, $r_\mathrm{u}$, d) gating off duration, $T_\mathrm{g}$, e) channel length, $l_{\mathrm{ch}}$, and f) gating outlet channel length, $l_{\mathrm{go}}$.}
        \label{Ap_as} 
     \end{figure*}

The second part of our analysis focuses on assessing the capability of our model to accurately represent the characteristics of the sampled signal under different scenarios. In Section \ref{sec:analytical model}, particularly in the optimization part, our main objective was determining the optimal range in which our analytical model yields accurate approximations. Additionally, we aimed to highlight the points of divergence between the proposed models and simulation results, effectively illustrating the areas where they differ from each other. The pulse amplitude plays a crucial role in communication system design, as it directly influences the SNR. Hence, we compare the pulse amplitude, $A_\text{p}$, obtained using \eqref{ap2_2}, with the numerically computed results, as shown in Fig. \ref{Ap_as}. As observed, the simple expression provided in \eqref{ap2_2} remarkably approximates the exact results and follows the trends as parameter values vary. The results demonstrate that nearly all parameters significantly impact the pulse amplitude, and the proposed analytical model effectively approximates the exact numerical results and their corresponding trends. Beyond a specific range, the model's accuracy in approximating the results diminishes. This divergence between the proposed approximate model and the simulation results is evident in Fig. \ref{Ap_as}(f).

\begin{figure*}
            \centering
     \begin{subfigure}[b]{0.32\textwidth}
         \centering
         \includegraphics[width=\textwidth]{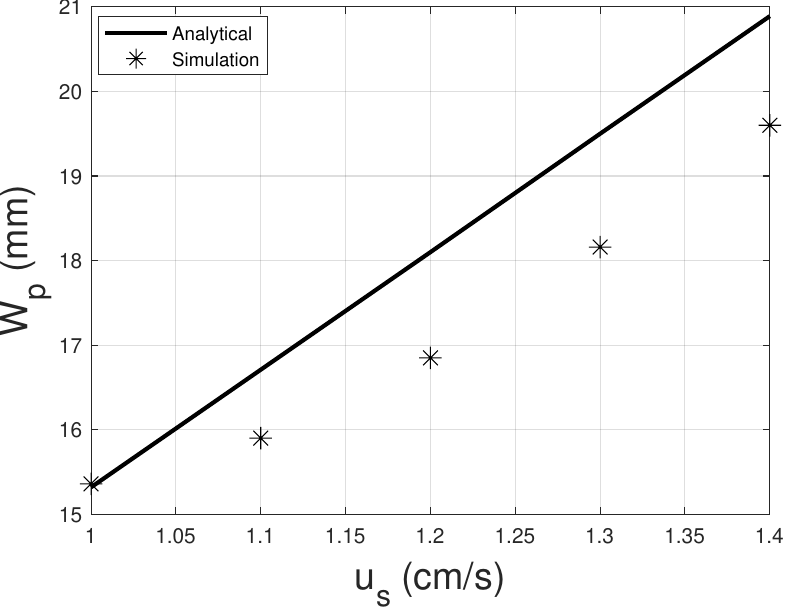}
         \caption{}
         \label{}
     \end{subfigure}
     \hfill
     \begin{subfigure}[b]{0.32\textwidth}
         \centering
         \includegraphics[width=\textwidth]{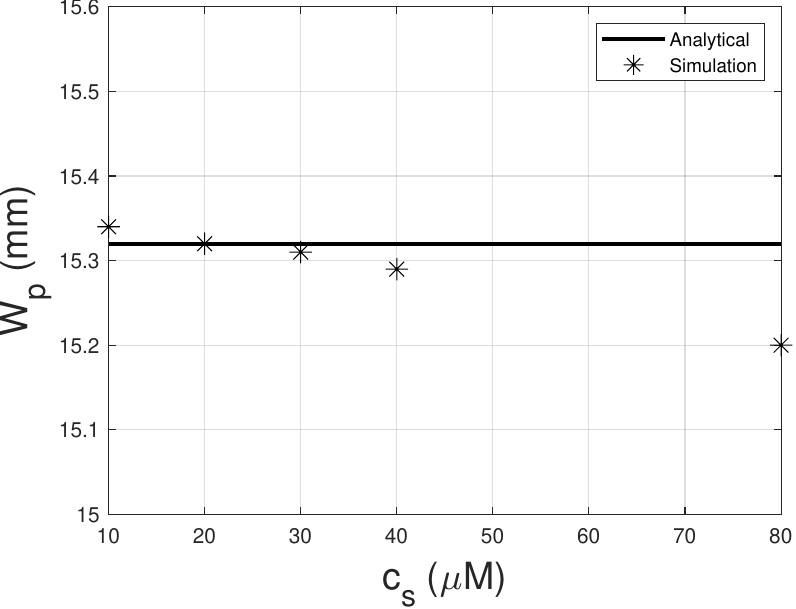}
         \caption{}
         \label{}
     \end{subfigure}
        \hfill
             \begin{subfigure}[b]{0.32\textwidth}
         \centering
         \includegraphics[width=\textwidth]{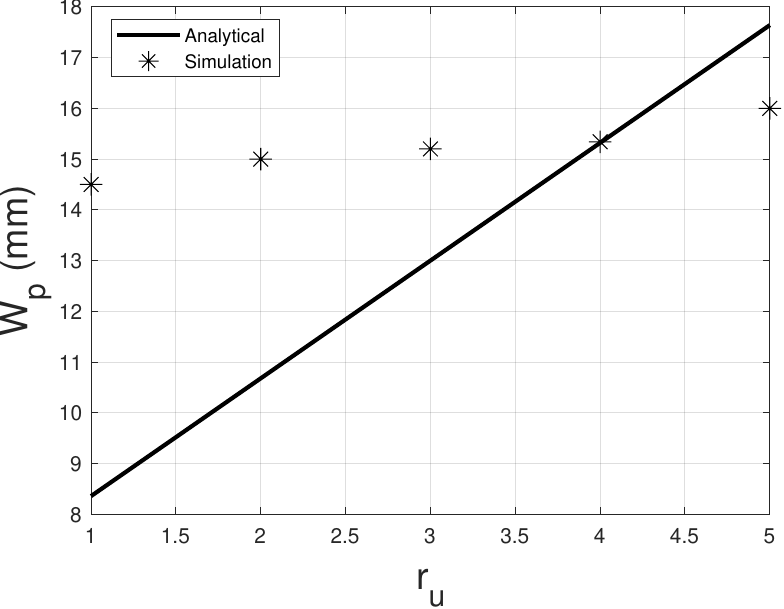}
         \caption{}
         \label{}
     \end{subfigure}

         \begin{subfigure}[b]{0.32\textwidth}
         \centering
         \includegraphics[width=\textwidth]{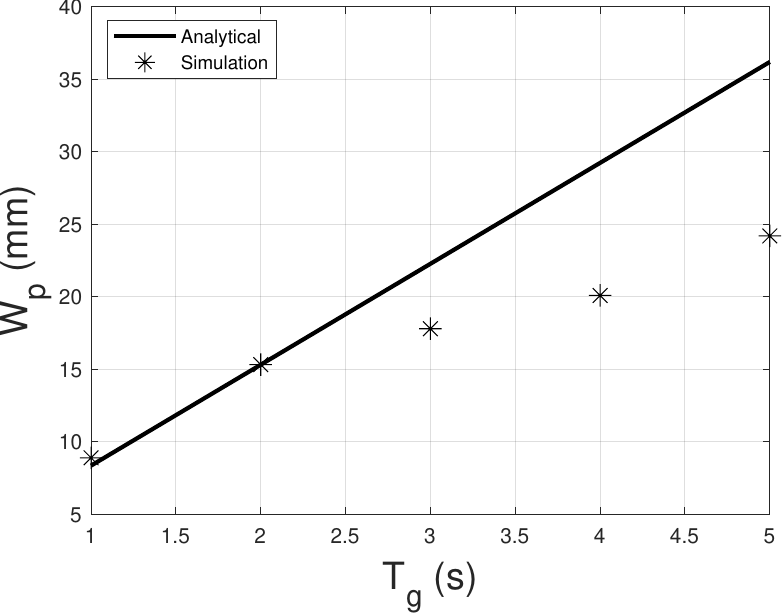}
         \caption{}
         \label{}
     \end{subfigure}
        \hfill
             \begin{subfigure}[b]{0.32\textwidth}
         \centering
         \includegraphics[width=\textwidth]{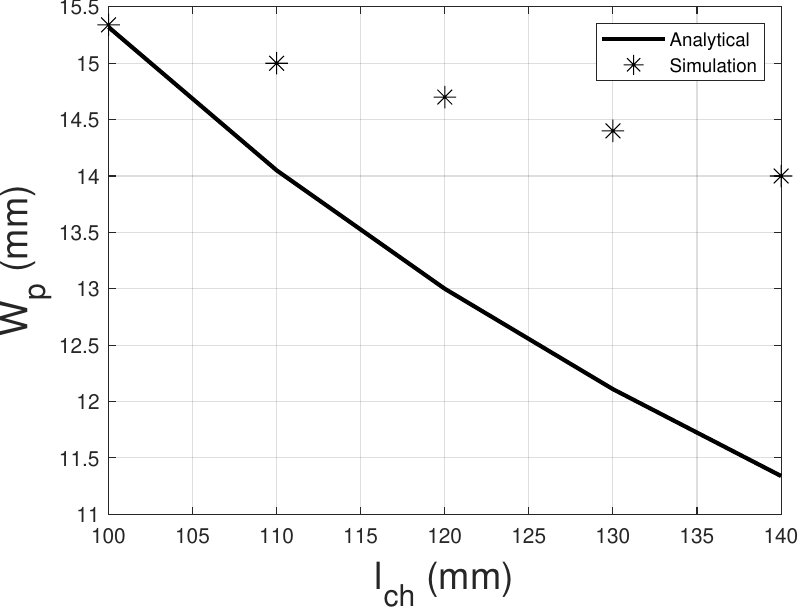}
         \caption{}
         \label{}
     \end{subfigure}
        \hfill
     \begin{subfigure}[b]{0.32\textwidth}
         \centering
         \includegraphics[width=\textwidth]{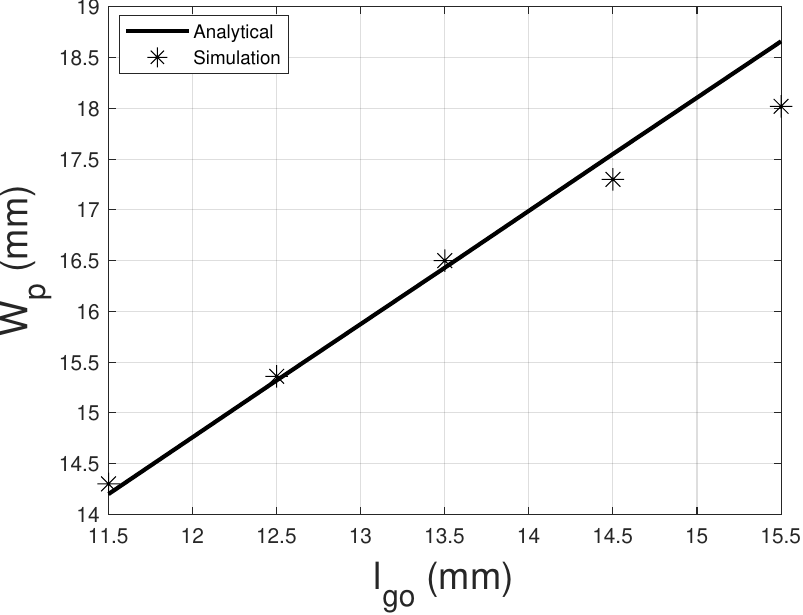}
         \caption{}
         \label{}
     \end{subfigure}
           \caption{\textbf{Pulse width, $W_\mathrm{p}$, for varying system input parameters:} a) flow velocity, $u_\mathrm{s}$, b) input concentration, $c_\mathrm{s}$, c) ratio between velocity of supply flow and gating flow, $r_\mathrm{u}$, d) gating off duration, $T_\mathrm{g}$, e) channel length, $l_{\mathrm{ch}}$, and f) gating outlet channel length, $l_{\mathrm{go}}$.}
        \label{Wp_as} 
     \end{figure*}

The analyses are repeated for the pulse width, $W_\mathrm{p}$, a crucial parameter directly associated with the extent of ISI and achievable bandwidth. Fig. \ref{Wp_as} illustrates the comparison between analytical results obtained using \eqref{wp2_2} and numerical calculations for the sampled pulse width. The proposed model effectively captures the characteristic trends observed under varying conditions. Notably, the pulse width remains unaffected by the initial concentration of ligands, $c_\mathrm{s}$. However, all other input parameters have an impact on the pulse width.

\begin{figure*}
         \centering

     \begin{subfigure}[b]{0.32\textwidth}
         \centering
         \includegraphics[width=\textwidth]{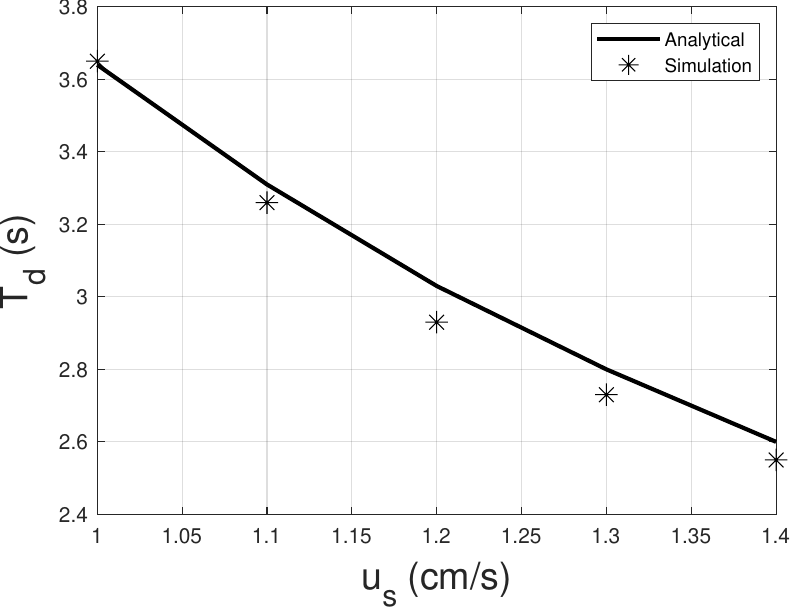}
         \caption{}
         \label{}
     \end{subfigure}
\hfill
     \begin{subfigure}[b]{0.32\textwidth}
         \centering
         \includegraphics[width=\textwidth]{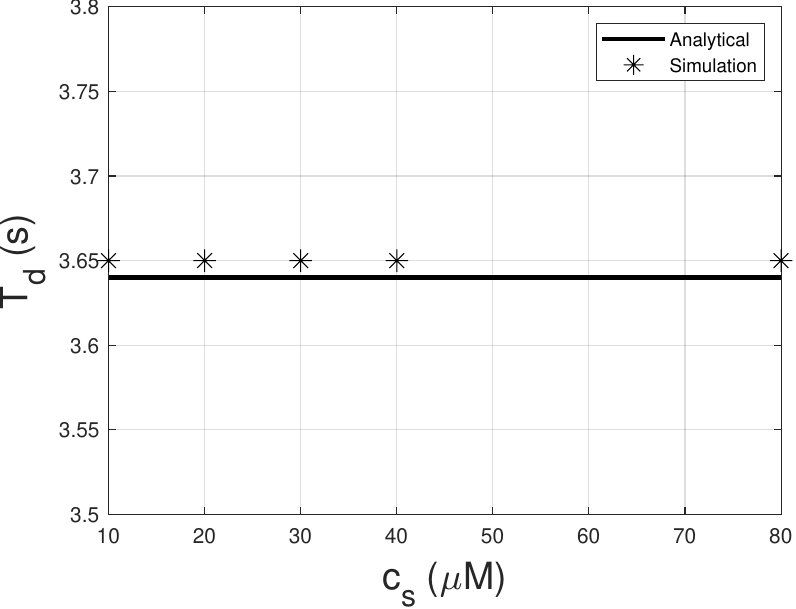}
         \caption{}
         \label{}
     \end{subfigure}
\hfill
     \begin{subfigure}[b]{0.32\textwidth}
         \centering
         \includegraphics[width=\textwidth]{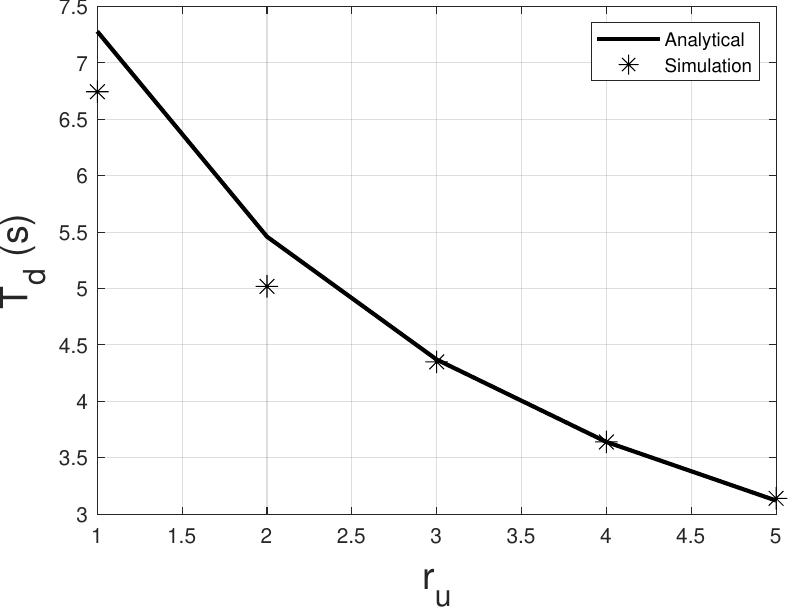}
         \caption{}
         \label{}
     \end{subfigure}

     \begin{subfigure}[b]{0.32\textwidth}
         \centering
         \includegraphics[width=\textwidth]{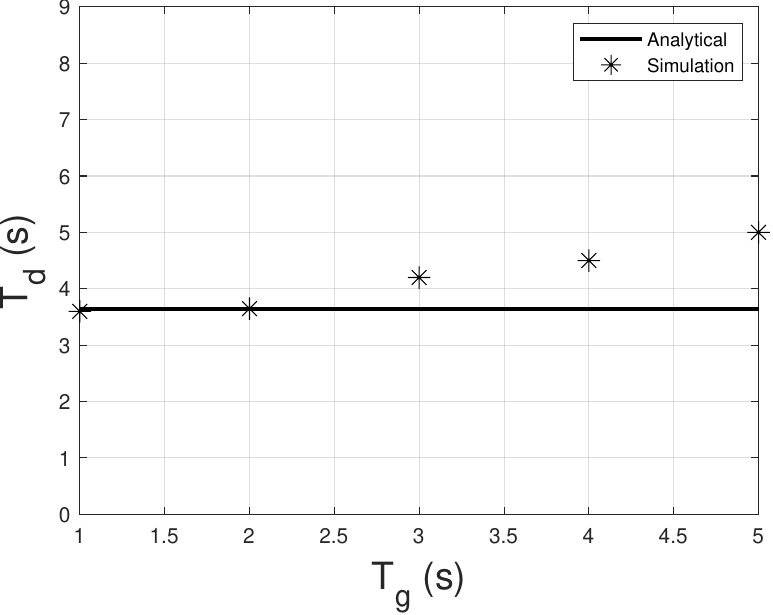}
         \caption{}
         \label{}
     \end{subfigure}
\hfill
     \begin{subfigure}[b]{0.32\textwidth}
         \centering
         \includegraphics[width=\textwidth]{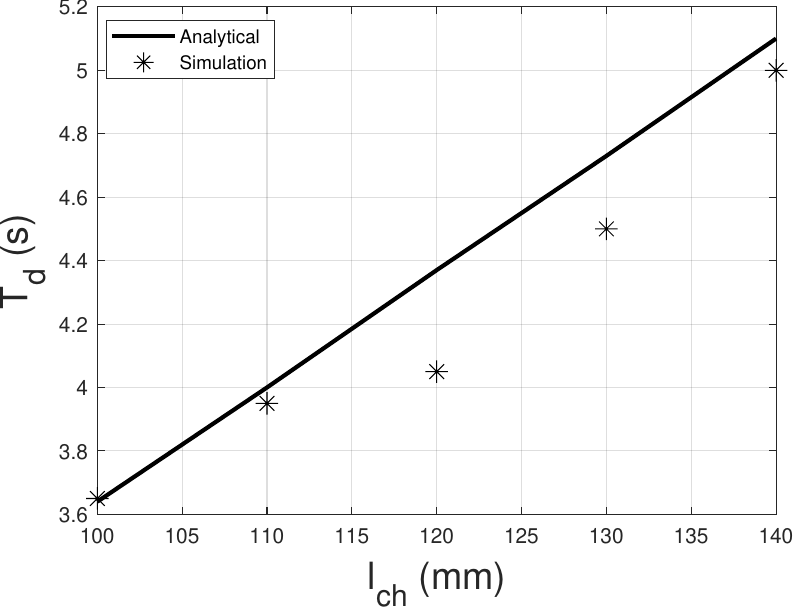}
         \caption{}
         \label{}
     \end{subfigure}
     \hfill
          \begin{subfigure}[b]{0.32\textwidth}
         \centering
         \includegraphics[width=\textwidth]{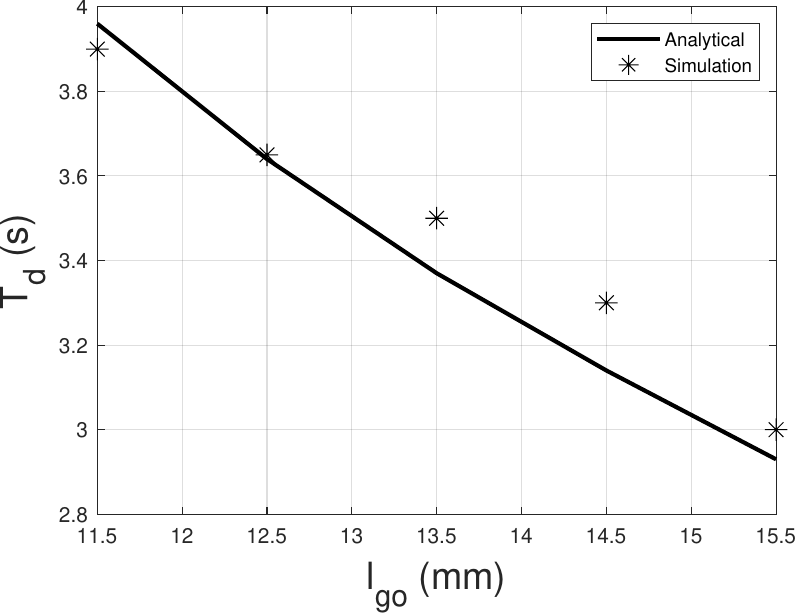}
         \caption{}
         \label{}
     \end{subfigure}

      \caption{\textbf{Pulse delay, $T_\mathrm{d}$, for varying system input parameters:} a) flow velocity, $u_\mathrm{s}$, b) input concentration, $c_\mathrm{s}$, c) ratio between velocity of supply flow and gating flow, $r_\mathrm{u}$, d) gating off duration, $T_\mathrm{g}$, e) channel length, $l_{\mathrm{ch}}$, and f) gating outlet channel length, $l_{\mathrm{go}}$.}
        \label{Td_as} 
\end{figure*}

The last set of analyses is carried out for the pulse delay, $T_\mathrm{d}$. As depicted in Fig. \ref{Td_as}, the analytical results obtained using \eqref{td} are compared with the simulation results, and they exhibit a high degree of agreement. Except for $T_\mathrm{g}$ and $c_\mathrm{s}$, all input parameters exert a considerable effect on the pulse delay. Our model, being simple and practical, does not necessitate computationally expensive numerical methods. It accurately captures design tradeoffs and proves valuable in designing efficient and reliable microfluidic MC systems prior to final implementation.

\section{Conclusion}
\label{section:conclusion}
In this work, we present an analytical model for a microfluidic MC transmitter that uses the hydrodynamic gating technique. We provide analytical expressions for the transmitted pulse's amplitude, width, and delay to aid in the optimization of microfluidic systems from a communication perspective. Our model was found to be approximately accurate based on a comparison of numerical results from the simulation and the model. In the next phase of this work, we will integrate a biosensor-based receiver into the model to optimize it for communication and experimentally test the model. Additionally, we will investigate new, low-complexity MC modulation techniques compatible with hydrodynamic gating.

\bibliographystyle{IEEEtran}

\bibliography{references}

\end{document}